# Migration and Educational Assortative Mating in India: How Geographic Mobility Shapes Marriage Markets

**Minali Grover**

Indian Institute of Management Indore, India

Jindal Global Business School, O.P Jindal Global Univeristy, Haryana, India

**Ajay Sharma**

Indian Institute of Management Indore, India

Global Labor Organization, Essen, Germany

**Declaration of Interest Statement:** We know of no conflicts of interest associated with this publication, and there has been no financial support for this work that could have influenced its outcome. As the corresponding author, I confirm that the manuscript has been read and approved for submission by all the named authors.

**Data availability:** The data (Periodic Labour Force Survey: 2020-21) used in the study is available at https://microdata.gov.in/nada43/index.php/catalog/PLFS.

**Funding Declaration:** This research received no specific grant from any funding agency or institution.

**Corresponding Address:** Minali Grover, Indian Institute of Management Indore, Prabandh Shikhar, Rau-Pithampur Road, Indore, Madhya Pradesh, India-453556; e-mail: f21minalig@iimidr.ac.in

Ajay Sharma, J-206, Academic Block, Indian Institute of Management Indore, Prabandh Shikhar, Rau-Pithampur Road, Indore, Madhya Pradesh, India-453556; e-mail: ajays@iimidr.ac.in, ajaysharma87@gmail.com

# Migration and Educational Assortative Mating in India: How Geographic Mobility Shapes Marriage Markets

*Abstract*

*This paper examines how internal migration influences educational assortative mating patterns in India using Periodic Labour Force Survey data (2020-21). We analyze the association of migrant status and type of assortative mating, that is whether migrants are more likely to engage in homogamous (similar education) or heterogamous (different education) marriages compared to non-migrants. Results show migrants are significantly more likely to form educationally heterogamous marriages, with urban-to-urban migrants particularly prone to hypogamy (marrying higher-educated partners). These findings are validated using instrumental variables including crime rates, migrant networks, and unemployment rates. Family structure and marriage pool composition emerge as key mechanisms driving educational heterogamy among migrants, suggesting migration fundamentally alters marital formation preferences away from traditional educational homogamy patterns.*

**Migration and Educational Assortative Mating in India: How Geographic Mobility Shapes Marriage Markets**

1. Introduction

The spatial distribution of individuals has always been of interest to researchers. In labour economics, how individuals and workers are spatially located determines their access to education, job opportunities, and social networks. Thus, this influences their lifelong career, income, social trajectories, and inter-generational outcomes (Chetty & Hendren, 2018). While in labour economics, spatial distribution dictates the movement of firms and jobs along with workers (Kain, 1968), in sociology and demography, the geographical location is closely tied to social relations and family formation (South & Crowder, 1999). Alongside the spatial distribution, individuals' mobility determines their labour and social outcomes. Although researchers have extensively explored the implications of mobility and migration on labour market matching[1], relatively little attention has been given to how such mobility influences social outcomes, particularly marital matching or assortative mating patterns[2]. Apart from determining family formation, assortative mating also shapes how resources are shared between and within families (Boertien & Permanyer, 2019; Schwartz, 2010; Schwartz, 2013) in a society. Moreover, impacting the inter-generational resource allocation in the household.[3] Thus also affecting the future populations. Whilst the extant literature on assortative mating focuses extensively on ascribed traits such as religion, race, and ethnicity (see review, Kalmijn, 1993), we focus on achieved trait, that is, educational level, which is relatively less explored.

We attempt to study how migration impacts educational assortative mating. Specifically, whether the decision to migrate improves matching in the marriage market, that is, leads to educational homogamy (wife's and husband's education are equal) among migrants as compared to non-migrants in India. A higher rate of homogamy reflects the degree to which "social structures and networks are closed to outsiders" (Blossfled, 2009). In other words, homogamy indicates how individuals are restricted and closed towards their traits. However, in recent decades, there has been a decrease in educational homogamy in developed countries (Blossfled, 2009; Schwartz, 2013).

---

[1] Past studies have looked at how migration decreases the mismatch in the labour market by addressing labour shortages, skill matching, balancing regional disparities, and assimilating cultural and economic systems (Jia et al., 2024; Lucas, 2016; Preston & McLafferty, 1999)

[2] Assortative mating refers to the "non-random matching of individuals into relationships" (Schwartz, 2013). Meanwhile, positive (negative) assortative mating exists when individuals with similar (dissimilar) traits are more likely to mate with each other.

[3] For instance, Beck and Gonzalez-Sancho (2009) demonstrate that equal education among partners (educational homogamy) is associated with children's school outcomes.

Theoretically, modernization and industrialization theory (Smits et al., 1999) predicts a fall in educational homogamy primarily for two factors.

First, with industrialization, as the economy expands, individuals become more mobile, increasing interaction among people from different demographic and socioeconomic groups. Hence, the geographical boundary of spousal selection is expanded. This is especially relevant in regions where the marriage pool is constrained by religion, caste, ethnicity, and occupation. For instance, the low sex ratio in Uttar Pradesh (a state in India) has resulted in the movement of brides from areas with relatively better sex ratios, such as West Bengal (Kaur, 2012). Similarly, in the international migration context, Thai (2005) provides an example of Vietnamese natives marrying U.S. low-wage earners. This phenomenon is also referred to as spatial hypergamy, where marrying a migrant is expected to render an upgrade to one's social status. In the context of internal migration, the Chinese household registration system (hukou) has enhanced the role of location as a feature of spousal selection, and marriage migration has become a way to move to a favourable location (Fan & Li., 2012; Mu & Yeuny, 2019; Sun et al., 2021). In a nutshell, modernization coupled with demographical imbalances alters the regional assortative mating patterns, which could also lead to a move away from the traditional norm of homogamy (mating across individuals with similar cultural and socioeconomic features) to heterogamy. We look at whether this widening geographical pool impacts the educational homogamy pattern and whether it is valid among internal migration in the context of developing countries.

Second, industrialization and modernization theory also suggest that as the economy progresses, the influence of parents' control on marriage decisions decreases (Blossfled, 2009). Along with it, migration could further decrease the influence of third-party members[4] which includes family, kinship, and community networks as individuals move out of their origin place. It also alters the social and cultural boundaries in which an individual functions, thereby increasing the likelihood of exogamy (marrying outside the community). Empirically, Esteve & Bueno (2012) examine the international migrants from Morocco to Spain and observe that the migrants are more likely to be in an exogamous marriage after 8 years of residence in the destination country. This also relates to the shift from the traditional assortative mating, where sorting occurred based on ascribed traits (which includes religion, caste, ethnicity, family background, etc.) but now has shifted to achieved traits (education and occupation)[5] (Kalmijn, 1991; Schwartz, 2013). Past literature has verified this shift extensively both in the context of developed countries (Blossfeld, 2009; Choi & Mare, 2012; Qian & Lichter, 2007) and developing countries (Lin et al., 2020; Sun et al., 2021). Overall, education is not only a prime signal in the labour market but has evolved to become a significant characteristic in the marriage market (see review, Blossfled, 2009). For migrants, the parent's control and social norms could change significantly; thus, how these factors impact their marriage preferences is less explored, especially for internal migrants.

---

[4] Variations in assortative mating patterns over a period of time have been pushed by modernization factors such as urbanization, mobility, and economic development, which have altered mate selection behaviours (Schwartz, 2012).
[5] Under the modernization theory, another hypothesis posits that preference for achieved traits rises as the importance of education increases and then falls as love becomes the driving force of spousal selection. Thus, following an inverted U-shape (Smits et al., 1998). It is also known as the "romantic love hypothesis".

Coming to educational assortative mating (hereafter, EAM) patterns, traditionally, educational hypergamy (the wife's education is less than the husband's education) persists; however, studies have observed a shift towards educational hypogamy (the wife's education more than the husband's education) in recent decades. This is also termed the "end of hypergamy" (Esteve et al., 2012). With increasing levels of women's education, the pattern in developing countries such as India corroborates with the global experience of an increase in educational hypogamy (Laha & Pradhan, 2024; Sarkar, 2020). However, unlike developed countries, India's educational hypogamy is not coupled with reverse gender norms but is a sum of traditional marriage norms of patrilocal exogamy, consanguineous marriage, and arranged marriages (Lin et al., 2020). Moreover, female migration is primarily driven by marriage migration (88 percent) in India (PLFS 2020-21), thereby dictating women's spatial mobility. Keeping in mind India's marriage migration, we attempt to explore the decision to migrate on type of EAM (homogamous and heterogamous marriage). Within heterogamous, we look for the existence of hypogamy and hypergamy. Further, considering the heterogeneity in the sample, we look at demographic characteristics (rural and urban), couple migration status (both migrant, husband-migrant, wife-migrant, and both migrant), spatial dimensions consisting of migration stream (rural-rural, rural-urban, urban-urban, and urban-rural) and migration type (intra-district, inter-district, and inter-state), and reason of migration (employment, marriage, and others). The OLS method provides the association between migrant status and EAM. We complement our analysis with instrumental variable strategy. We use migrant network, crime, and unemployment rate as instruments which impacts migration but not EAM[6]. The IV estimates are local average treatment effects (LATE) as they capture individuals whose migration decisions are impacted by economic (unemployment), network dependent (migrant network), and safely conscious (crime) constraints. Moreover, we complement instruments with heteroskedasticity-based instruments using Lewbel IVs (2012) to overcome any weak instrument problem. The main empirical challenge is to establish causality between migration and EAM. Since, migration and marriage decisions are often jointly determined, it is difficult to establish the temporal ordering in a cross-sectional dataset.

---

[6] We discuss about the exclusion restrictions and IV strategy in the sub-section in the Empirical Methodology section.

Therefore, the results are to be interpreted as association between migration status and EAM. Additionally, the instrumental variable strategy addresses some selection concerns but does not fully eliminate the simultaneity or establish causality. Lastly, to uncover the potential mechanisms, we look for the impact of third-party influence (nuclear versus joint family) after controlling for the sample selection bias and the sex ratio differential between the origin and destination states of a migrant (geographical pool).

To do so, we use India's latest nationally representative dataset, the Periodic Labor Force Survey (PLFS) 2020-21, which has particulars related to migration, employment, human capital, and job-market characteristics. Additionally, we include controls and instrumental variables from the National Crime Records Bureau, PLFS (2019-20), and Census (2011) to complement our analysis. We found an overall decrease in educational homogamy, which is replaced by an increase in educational hypogamy as we move from older to younger cohorts. The pattern corroborates with the findings observed by related studies (Lin et al., 2020; Roychowdhury & Dhamija, 2024; Sarkar, 2022) for India. Additionally, among migrants, a rise of educational heterogamy is even starker as compared to non-migrants. Thus, some additional factors are driving the EAM patterns for migrants. Using logit regression analysis, the findings suggest migrants are less likely to be in a homogamous marriage and are even less likely in rural areas. Additionally, we complement our analysis with the instrumental variable analysis using migrant network, the unemployment rate in rural and urban areas, and log of crime along with heteroskedasticity-based instruments (Lewbel, 2012) that provide robustness to our findings. However, we are unable to establish a causal claim between EAM and migrant status due to lack of timing and duration of migration in our data. Within heterogamy, using multi-logit analysis, we look at hypogamy and hypergamy unions and found that female migrants have higher odds of marrying up educationally as compared to homogamy. Thus providing evidence for association between educational hypogamy among migrants in India. To identify the driving mechanisms, we analyse the impact of co-habiting in-laws to look for third-party influence in marriage preference and relative sex differentials of migrants' origin and destination states to dissect the potential geographical pool. The result indicates that migrants in nuclear families have relatively lower odds of being in an educationally homogamous union, and female migrants have higher odds of educational hypogamy in nuclear households. In line with the extant literature, migration reduces the influence of community, kinship, and family on marriage preferences, which leads to a shift from ascribed to achieved traits. Moreover, migration broadens the geographical scope, which increases the channels of meeting potential spouses outside one's kin. This is also observed when we look at the spatial characteristics of a migrant; as the migrant's distance increases from intra-district to inter-district, the likelihood of homogamy further reduces.

The paper contributes to two broad strands of literature. The first strand of literature is concerned with migration and assortative mating. As the migration decision coincides with the time of marriage decision or having children, the migration pattern could dictate the selection of marital formation. Studies exploring assortative mating primarily examine international migration and are concentrated in developed countries (see for example, Choi & Mare, 2012; Dziadula & Zavodny, 2024; Potarco & Bernadi, 2024). For instance, Choi & Mare (2012) look at the migrants from Mexico to the U.S. and find that migrants are more likely to be in an educational heterogamous marriage. Within heterogamous unions, migrants have a higher likelihood of "marrying up" in educational level than non-migrants. Further, validating the status-exchange theory, Potarco and Bernadi (2024) found that highly educated immigrants exchange their migration status by marrying down in terms of education with the Swiss natives. Thus, trading off their foreign status by compensating with the lower education of their native spouse.

A higher educational level is also linked with exogamy (marrying outside their nationality) for both males and females (Esteve & Bueno, 2012), thus indicating that education is linked with weakening social and cultural boundaries among immigrants. While these studies are explored for international migrants, the internal migration dimension in EAM is relatively scarce. A handful of studies concerning China and its internal migration highlight that internal migration increases the match along the achievement traits rather than ascribed traits (Hu, 2024; Sun et al., 2021). Moreover, educated internal migrants are more likely to be in an exogamous marriage (Mu & Yeung, 2019), exhibiting evidence of weakening third-party influence as individuals migrate. Lastly, using mixed-method research, Mu & Yeuny (2019) indicate that female migrants tend to "marry up" primarily. To sum up, studies on EAM and internal migration portray evidence towards non-traditional marital formation, such as educational hypogamy and exogamy. In this paper, by considering internal migration in India, we add to the current literature on how migration impacts mating patterns and the potential mechanism through which marriage preferences are altered.

The second strand of literature explores the educational assortative mating patterns in India. Overall, the studies can be separated into two broad themes: the first set of literature examines the influence of EAM on women-related dimensions, and the second set of studies validates the status-exchange theory. The impact of educational hypogamy is explored in terms of female employment (Roychowdhury & Dhamija, 2024), dowry (Goel & Barua, 2023), and domestic violence (Roychowdhury & Dhamija, 2022) for India. Overall, women in hypogamous marriage are more likely to experience domestic violence (Roychowdhury & Dhamija, 2022) and has lower likelihood to participate in the labour force (Roychowdhury & Dhamija, 2024). In a patriarchal society like India, educational hypogamous marriage deviates from the traditional marriage system thus, could lead to backlash effect from the groom's family in terms of marital conflicts, loss of autonomy, and marital stress (Roychowdhury & Dhamija, 2024). On the contrary, Goel and Barua (2023) explore the impact of the wife's educational levels on dowry and conclude that they are inversely related. In other words, the bride's (groom's) higher education entails a lower (higher) dowry, thus depicting a substitutionary relationship. Another set of studies of EAM in India examines whether a trade-off exists between higher education and other marriage market qualities. Following the status exchange theory[7] (Davis, 1941), Lin et al. (2020) and Sarkar (2022) emphasize that educational hypogamy is associated with caste and socioeconomic hypergamy. In other words, marrying down by wife in terms of education is compensated by marrying up in socioeconomic class, caste, or occupation. Overall, the recent literature on EAM in India overlooks the impact of migration on marriage-matching. Since internal migrants comprise 28.9 per cent of the total population (rural, 26.5 per cent and urban, 34.9 per cent) (Chandrasekhar & Sharma, 2022), they cover approximately one-fourth of the population. As a result, how their social outcomes and family formation are different from those of non-migrants could highlight whether the mobility of individuals impacts EAM.

To briefly summarize, the paper contributes to the existing literature in three broad ways. First, this study focuses on linking migration and EAM using the nationally representative data for India. Second, as a robustness, we employ instrumental variable analysis along with establishing potential mechanisms using family structure and sex-ratio origin/destination differentials. Lastly, while extant literature looks at the international migration, we look at internal migration and explore the heterogeneity in terms of couple migrant status, reasons of migration, and spatial factors (migration type and stream).

The following section explains data sources and highlights descriptive statistics. Section three discusses the empirical methodology for the ordinary least squares (OLS) and instrumental variable approach. Section four discusses results, heterogeneity analysis, and potential mechanisms. The paper closes with a discussion in section five.

**2. Data and Descriptive Statistics**

*2.1 Data Sources*

---

[7] The status-exchange theory suggests individuals pair up mutually complimentary traits with each other. For instance, a lower caste of a spouse might be compensated by a higher socioeconomic status or vice versa. The theory has been extensively explored in the context of both developed countries (see, for example, Celikaksoy et al., 2006; Kalmijn, 1994; Potarca & Bernadi, 2017) and developing countries (Lin et al., 2020; Sarkar, 2022).

To ascertain the impact of the decision to migrate on educational assortative mating (EAM), we employ the nationally representative dataset, the Periodic Labor Force Survey (PLFS) 2020-21 conducted by the National Statistical Office (NSO 2022) from July 2020 to June 2021. The survey comprises 1,00,344 households (4,10,818 individuals), out of which 55,389 households (2,36,279 individuals) reside in rural, and 44,955 households (1,74,539 individuals) reside in urban areas. The survey provides detailed information about demographic characteristics, human capital, employment particulars, industry and occupation classification, and wage details for all individuals in the household. Additionally, unlike the other rounds of the PLFS survey, the 2020-21 round also had migration particulars for the household members. Hence, this is the latest nationally representative survey providing migration information. The survey defines *a migrant* as an individual whose "last usual place of residence is different from the present place of enumeration", where the usual place of residence is a village/town where an individual has resided continuously for more than six months. The relationship to head and person serial number helped us identify the couples in the survey (annexure A.1 and table A.1 provides more details for the couple identification strategy). As per the requirement of the study, we limit our analysis to the working-age population, thereby including individuals above 15 years and below 59 years of age.

To supplement the analysis, we further use the latest round of Census 2011, National Crime Records Bureau (NCRB), and preceding round of PLFS survey (2019-20) for covariates and instrumental variables. Since 1971, NCRB has provided district-wise annual crime data for India; the statistics are based on crime reporting. We consider the incidence of crimes for the year 2020. For the analysis, we used NCRB data from the Centre for Economic Data and Analysis (CEDA)[8] which provides separate crime-wise files for districts in India. Next, using census data, we calculate the proportion of eligible males (15-24 years) to females (20-26 years) as a control variable. The migrant network used as an instrumental variable is also calculated using the Census. The unemployment rate from the preceding round of the PLFS survey (2019-20) is included as an instrument variable. The variables are discussed in detail in the empirical section (3.2).

*2.2 Educational Assortative Mating Patterns*

Before delving into the empirical methodology and regression estimates, this section summarises descriptive patterns concerning EAM and migration in India. As past studies highlight, there has been a consistent increase in heterogamous marriages in India, primarily in educational hypogamous marriages (Lin et al., 2020;

---

[8] The Economics Department of Ashoka University developed the CEDA database in 2020. It provides a comprehensive data portal on key socioeconomic indicators in easily accessible formats.

Roychowdhury & Dhamija, 2024). Table 1 highlights an approximately 8 per cent difference in educational homogamous marriages between older and younger age cohorts (45-59 and 15-24). Within heterogamous marriages, the gap between younger and older cohorts is larger for hypogamous marriages. Thereby indicating an increase in educational hypogamy among the younger cohorts. While the majority of marriages are still hypergamy where the husband is more educated than the wife, however, this aligns with Becker's (1983) traditional model, which points towards educational hypergamy among developing and less-developed countries, especially among older cohorts (Blossfled, 2009). The transition tables (appendix table A.4) exhibit a similar pattern where migrants constitute a higher proportion of heterogamous marriages, especially at higher educational levels. A similar pattern is observed by Lin et al. (2020) and Sarkar (2022), showing that educational hypogamy has increased from 10 percent to 30 percent in the last four decades. Likewise, Roychowdhury and Dhamija (2024) state that educational hypogamy rose from 7 percent to 24.1 percent from 1992 to 2021 in rural areas. Upon bifurcating the sample for rural and urban areas (presented in annexure table A.2), the hypogamy pattern indicates that differences between younger and older cohorts are stark in rural areas. While in rural areas, there is a shift from homogamy to heterogamy marriages from older to younger cohorts, in urban areas, the shift occurred among hypogamy and hypergamy marriages. In urban areas, the prime driver could be a steady rise in female educational levels.

<Table 1 here>

Coming to our interest variables, marriage patterns for migrants and non-migrants, we look at the distribution of EAM and age cohorts across migrant status (table 1, panels b and c). There are two pivotal observations. First, 38 percent of non-migrants (15-24 age group) are in educational homogamous marriages as compared to 34 percent of migrants. Second, within heterogamous marriages, hypogamous marriages comprise 28 percent of migrants as compared to 26 percent of non-migrants. However, the pattern holds primarily for the young cohort of individuals, which indicates a shift towards hypogamy.

Further, we bifurcate the couple migrant status into non-migrants, husband-migrant, wife-migrant, and both migrants in Table 2. We observed that the proportion of homogamous marriages is highest when both individuals are non-migrants. This indicates that non-migrants might have socially closed boundaries restricting them from joining homogamous unions. Within heterogamous unions, hypergamy marriages are almost twice (41 per cent) that of hypogamy marriages (19 per cent), indicating a trend of women "marrying up" educationally. Lastly, among heterogamous marriages, hypogamous (hypergamous) marriages are highest when the husband (wife) is a migrant.

<Table 2 here>

Overall, the EAM patterns discussed above provide a basis for further analysis, depicting that the proportion of migrants is higher in heterogamous marriages.

3. **Empirical Methodology**

This section discusses the empirical methodology. The first sub-section outlines the ordinary least squares (OLS) method followed by explaining the instrumental variable strategy in the second sub-section.

*3.1 Ordinary Least Squares*

To investigate how educational assortative matching (EAM) varies with migrant status, we estimate the following regression equation:

$$AM_i = \beta_o + \beta_1 M_i + \beta_2 X_i + \epsilon_i \tag{1}$$

where $AM_i$ refer to assortative matching, which is equal to 1 if individual $i$ has an equal educational level[9] (homogamous marriage union) with their spouse, and zero otherwise (heterogamous marriage union). $M_i$ indicates whether an individual is a migrant. $X_i$ incorporates set of control variables which includes educational levels (illiterate, primary, middle, secondary, higher-secondary, and graduate and above), age-cohorts (15-24, 25-34, 35-44, and 45-59), gender (male and female), sector (rural and urban), religion (Hindu, Muslim, and others), social group (schedule caste, schedule tribe, other backward classes, and others), and log of monthly per capita expenditure (MPCE). We further include dummies for 36 states and union territories. Lastly, to control for an eligible pool of males and females, we include the ratio of males (20-29 years) to females (15-24) at the district level for rural and urban areas separately as a covariate (Lin et al., 2020). Since the dependent variable is a binary outcome, we apply logit regression and restrict our sample to between 15 and 59 years old. The standard errors are clustered at the first-stage unit[10] (FSU), where sampling is conducted which is villages for the rural areas and locality/ward for the urban areas (Batra & Sharma, 2025). The summary statistics for the variables are presented in the annexure (table A.3). While for the summary statistics, we employ sampling weights, for the regression estimates, weights are not used. As a robustness, we verify the main regression results using sampling weights also (presented in the annexure table A.20).

---

[9] The results are robust after considering years of schooling instead of educational levels as the variable for educational assortative matching. For brevity, the findings are not reported and are available on request.
[10] The first-stage unit (FSU) is the basic geographical unit where sampling for the survey is conducted. Hence, clustering at the FSU level ensures standard errors are not overestimated.

PLFS provides information regarding migrant characteristics. Thus, we further explore the heterogeneities for migrants based on migration reasons (employment, marriage, and other), migrant type (intra-district, inter-district, and inter-state), and migration stream (rural-rural, rural-urban, urban-urban, and urban-rural). Furthermore, to accommodate spouse migrant status, we look at couple migration status.[11] (both non-migrant, husband migrant, wife migrant, and both non-migrant). We complement our analysis by establishing a potential mechanism; (i) type of family structure, and (ii) geographical pool determined by sex ratio differentials among origin and destination states of a migrant (discussed more in section 4.3).

*3.2 Instrumental Variable Estimation*

The OLS estimation strategy provides a relationship between migrant and educational assortative matching (EAM). However, it fails to consider the endogeneity of the prime variable, migrant status. Therefore, a causal relationship between migration and EAM is not estimated. We employ the instrumental variable (IV) approach as a robustness exercise. It is important to note that, given the nature of the data, we cannot establish causality between migration decision and assortative mating. Therefore, the results are to be interpreted as an association between migrant status and EAM. A variable is a valid instrument if it meets the following criteria. First, the variable should be correlated with the endogenous variable. Second, it should be uncorrelated with the error term. Third, it should not have any direct impact on the dependent variable.

Keeping in mind the above criteria, we introduce three instrumental variables. The instrumental variable should impact migration without having any direct impact on EAM. We consider the migrant network, unemployment rate, and crime as our instrumental variables. The migrant network is characterized as one of the critical factors for the decision to migrate (De Haas, 2010). Migrant networks reduce the costs and risks of movement at the destination and increase the likelihood of employment (Sharma & Das, 2018), thus raising the chances of migration. Therefore, the coefficient of the migrant network should be positively related to migration. Using the latest round of Census (2011), we consider the log of total migrants, which is the sum of migrants in rural and urban areas at the district level, as the indicator for the migrant network. By considering a lagged instrumental variable of approximately a decade, we attempt to ensure that migrant network does not directly influence the EAM at the time period of the data used (2020-21). While the 2011 migrant network determines where people migrated, the current EAM patterns are more likely to be current migration patterns. We plot migrant network and share of migrants to present their relationship in the annexure (figure A.1). The graph indicates, as migrant network increases, the share of migrants also rises. Alternatively, we also use another version of migrant network which weights the stock of migrant network with years of experience[12] for the robustness of the results. While for the main analysis we use stock of migrant network as instruments, we present the results for the main regression using alternate instrument of migrant network experience and migrant fractionalisation index[13] (based on Herfindahl-Hirshman index) (Suedekum et al., 2014) in the annexure (table A.19).

---

[11] Choi and Mare (2012) incorporate a couple migration status for international migrants from the U.S.A. to Mexico to determine assortative mating patterns.

Next, we consider the unemployment rate at the district level as another instrumental variable. Unemployment at the origin place could lead to migration to places where there are more employment opportunities (Greenwood, 1997; Eggert et al., 2010). Thus, unemployment in a region acts as a "push" factor for migrating to areas with lower unemployment rates and better economic prospects. We calculate the unemployment rate at the NSS region for rural and urban areas as our instrument using the preceding round of the Periodic Labor Force Survey (2019-2020). However, since PLFS provides migration particulars at the destination region, we would expect unemployment to be inversely proportional to the migrant status. In other words, the likelihood of migration in regions with a low unemployment rate should be higher. Using a lagged instrument could partially ensure that instrument does not impact EAM directly and only through migration. While in the long-run, studies show that how EAM could impact inequality through polarisation (Boertien & Permanyer, 2019; Schwartz, 2010) however, it is unlikely that, the EAM pattern alters as a result of changes in the unemployment in the short run. Alternatively, we also use unemployment rate from the EUS 2011-12 survey for robustness. The results are presented in the annexure table A.22.

Lastly, we consider crime at a district level as a factor affecting out-migration from a district. There are two channels through which crime affects migration. First, crime against women in a locality increases the likelihood of early marriage for females (Sarkar, 2024). Thus, raising the chances of out-migration for females as a result of marriage[14]. Second, higher crime in a district decreases the likelihood of out-migration of male members in a household where daughters are present. Bhat and Thakur (2024) portray evidence of an association between having more daughters and less odds of adult male members migrating. Therefore, local crime against women is associated with migration for both males and females. Using the National Crime Records Bureau (NCRB) and CEDA database (2020), we calculate the log of crime, which is the sum of violent and non-violent crimes (Mahesh, 2020) for the year 2020. We briefly discuss the first-stage results in the result section.

---

[12] Census (2011) provides duration of migration in four categories: less than 1, one to four years, five to nine years, and ten years and above.
[13] To measure the size effect of migrants coming from different state, we measure the share of migrants from specific state in each district and create a HHI index (Suedekum et al., 2012).
[14] Marriage-related migration constitutes around 85 percent of female migration in India (PLFS, 2020-21).

The use of multiple instruments strengthens the identification strategy by capturing different factors that impact migration. While we understand that these instruments could impact marriage, but it is unlikely that it could impact education matching. For instance, unemployment could impact both marriage and migration, but the chances that it would impact education matching are low. Therefore, we test for the exclusion restriction, we run a separate regression wherein we estimate the impact of instrument variables on AM. Ideally, there should be no impact of our instruments on non-migrants. The results are presented in the annexure table A.5. The absence of any relationship between education matching and instruments among non-migrants indicates instruments impact EAM through migration. Moreover, the estimates generated through IVs are local average treatment effect (LATE) that is, they capture estimates for individuals whose migration decision depends on these instruments. In other words, these compliers represent a subset of all individuals who are responsive to economic (unemployment), network dependent (migrant network), and safety conscious (crime) factors. To indicate the relevance of the instruments, the diagnostic tests showing the first-stage F-statistic, confirms strength of instruments.

Furthermore, to complement our IV model and overcome the weak instrument concern, we further implement Lewbel IV (2012)[15] which does not require traditional exclusion restrictions. A hybrid approach is provided by Lewbel (2012) wherein instrumental variables can be combined with heteroskedasticity-based instruments to rectify the weak instrument problem. Several authors have used a similar approach due to the absence of a valid instruments or combining heteroskedasticity based instruments with traditional instrumental variables (Bahl & Sharma, 2022; Dang & Rogers, 2016; Emran & Hou, 2013).

The pre-requisite to apply the Lewbel IV method is that the OLS model should have heteroskedastic errors. Using Breusch-Pagan/Cook-Weisberg tests[16], we identify the presence of heteroskedasticity in the OLS models for all the model specifications. Therefore, we reject the null hypothesis that the model has constant variances. Hence, Lewbel IV method can be applied. Moreover, since the instruments are generated using heteroskedasticity, they are uncorrelated with the error term. An important methodological consideration is that the traditional Lewbel IV method is applicable for the case with continuous endogenous variable (migrant status) however, our endogenous variable is binary in nature. As Lewbel (2018) discusses, heteroskedasticity based instruments with binary endogenous variables requires additional assumptions. Following Baum & Lewbel (2019), we conduct presence of heteroskedasticity in the errors of the first-stage regression using the Breush-Pagan test. The null hypothesis of homoskedasticity is strongly rejected (chi-square 3129.56, p-value 0.00 <0.01). We run some additional diagnostic tests for the IV models indicating instruments (under-identification tests) are relevant with sufficient high f-statistic value (annexure, table A.17).

---

[15] We implement the Lewbel IV (2012) using the statistical package ("ivreg2h") provided by Baum and Schaffer (2024) for STATA.
[16] The diagnostic tests are presented in the annexure (table A.17).

Using the two-stage least square approach, we employ instrumental variables and statistical-based instrumental variables to estimate the following first-stage equation:

$$M_i = \alpha_o + \alpha_1 Z_{1id} + \alpha_2 Z_{2id} + \alpha_3 Z_{3id} + \alpha_4 Z_{4id} + \alpha_5 S_{id} + \alpha_6 X_i + u_i \qquad (2)$$

where the dependent variable is the migrant status ($M_i$), $Z_{1id}$ refers to the log of crime, $Z_{2id}$ and $Z_{3id}$ indicate unemployment in rural and urban areas, respectively, and $Z_{4id}$ refers to the migrant network for individual I in district d. $S_{id}$ refers to a statistical-based instrumental variable, and $X_i$ refers to control variables discussed in the last sub-section. Lastly, $u_i$ refers to the error term. The first-stage results are presented in the annexure (table A.11 to A.15). In the second stage, we input the instrumental variables and estimate the following regression equation:

$$AM_i = \beta_o + \beta_1 \widehat{M_i} + \beta_2 X_i + \epsilon_i \qquad (3)$$

The variable interpretations are the same as equation (1). We emphasize that $\beta_1$ captures the association between migrant status and AM rather than causal estimates. The cross-sectional nature of the dataset restricts the temporal ordering of marriage and migration especially for women, as migration is induced from marriage in India. The IV estimates are discussed in the results section.

## 4. Results

We discuss the main results in the first sub-section, which is followed by an exploration of heterogeneity in the migrant's characteristics. In the last sub-section, we present two potential mechanisms.

*4.1 Main Results*

In Table 3, we present the relationship between migrant status and assortative mating, where the dependent variable is educational homogamy (equal education between wife and husband). The results indicate that a migrant is less likely to be in a homogamous marriage as compared to a non-migrant. Upon looking at sector-wise estimates separately, the likelihood of being in a homogamous marriage is significantly lower for individuals residing in rural areas as compared to urban areas. These results are based on OLS. To overcome the potential endogeneity, we incorporate three instrumental variables: migrant network, unemployment rate in rural and urban sectors, and log of crime, along with statistical-based instruments using the Lewbel IV method (Lewbel, 2012).

The first-stage results for the IV models (presented in the annexure, table A.11) suggest that the migrant network is positively related to migration. Meanwhile, higher crime and unemployment rates negatively impact migration in the destination region. This observation aligns with past studies indicating that migrant networks act as a pull factor for migration (Sharma & Das, 2018). In contrast, a lower unemployment rate at the destination region increases migration (Eggert et al., 2010). The presence of heteroskedasticity is a pre-requisite for using statistical-based instruments. The Breush-Pagan test statistics (presented in annexure table A.17) reject the null hypothesis of constant variance (homoskedasticity), confirming heteroskedastic errors. Thus, the model can implement the Lewbel IV (2012) method.

The IV estimates indicate a negative association between being a migrant and having an educational homogamous union. The IV estimates are slightly lower than the OLS results but are qualitatively similar. In order to verify the results are not primarily driven by the marriage-migrants, we compare non-migrants with non-marriage migrants. The results are presented in the annexure in table A.6. The estimates are in line with the main IV results indicating migrants are less likely to be in a homogamous union than non-migrants. While there are no comparable studies in the context of internal migration, EAM patterns in the international migration context portray similar evidence of heterogamy among migrant individuals (Choi & Mare, 2012). As discussed in the introduction, in line with modernization theory, migrants, by moving from their native place, widen the geographical pool and loosen social boundaries. Thereby, increasing the likelihood of being in a heterogamy union.

However, the result could be driven by the recent marriage cohort. Therefore, as a sensitivity check, we drop recent marriage cohort. Since the mean age of marriage in India is 19.2 (UNFPA, 2022) for women, we estimate the regression for individuals above 23 years to control for the recent marriages. As an additional check, we also report results for individuals aged 30 years and above for robustness. The results are presented in the annexure table A.7 indicates a negative likelihood of being in a homogamous for migrants which are in line with the main results.

<Table 3 here>

Within heterogamous marriages, using multinomial regression (refer to Table 4), we look for educational hypogamous (wife has more education than husband) and hypergamous (husband has more education than wife) marriages. Considering the base category as homogamous marriage, the estimates emphasize that migrants are more likely to be in hypergamous marriage as compared to non-migrants. Further, interacting migration status with gender, the likelihood of female migrants is weakly significant and positive for being in a hypogamous marriage. While for hypergamous marriage, the results are not sensitive based on gender. Similar to China's internal migration (Mu & Yeuny, 2019), wherein female migrants have higher odds of marrying up educationally, the Indian EAM exhibits a similar pattern. Though the results are weak for hypogamous unions among female migrants, it raises the question of whether the migration of females could be linked with a higher impact of modernization as compared to male migrants. On the contrary, Chatterjee and Desai (2022) highlighted how, in marriage migration, females exhibit characteristics of their "imagined" communities, which refer to their family's community norms. Thereby, even a weak significance for hypogamy could indicate drastic progress for females in India.

<Table 4 here>

Lastly, looking at the couple migration status (table 3, panel b) displays four main findings through instrumental variable estimates. First, irrespective of who the migrant is (husband or wife), a migrant is less likely to be in an educational homogamous marriage. Considering the couple-wise migration status further provides a strong result towards our primary finding that migrants move away from homogamy. Second, in terms of sector of residence, the lower odds of homogamy are primarily driven by urban areas regardless of whether the husband or wife is a migrant. Thirdly, the husband's migrant status has a substantially higher impact on heterogamy as compared to the wife's or couple's migrant status. Fourth, the wife's migrant status increases the odds of both hypogamy and hypergamy as compared to homogamy (table 4, Panel C). In a nutshell, in line with the theoretical background, migration impacts the social boundaries of an individual, altering their marriage preference. Thereby impacting their EAM patterns towards heterogamy. To summarize, we plot the main coefficients using the forest plot in figure 1, which highlights IV estimates are smaller in magnitude than OLS estimates and migrant status have significant impact on EAM irrespective of who migrates.

*4.2 Heterogeneity Analysis*

To explore the heterogeneity among migrants, we look at migrant's characteristics in terms of the reason for migration, migration stream, and migration type. The OLS and IV estimates for reasons of migration are presented in figure 2 (A). We plot the forest plot while the regression tables are presented in the annexure (table A.8 and A.9). Migration driven by marriage decreases the odds of a homogamous union, while there is an insignificant impact due to other reasons for migration. Both in the OLS and IV models, marriage migration significantly impacts EAM in urban areas. Since women constitute 98 percent of the total marriage migrants, we do a sub-sample analysis based on gender (annexure table A.10). The IV estimates suggest female marriage migrants and male employment-related migrants have significantly lower odds of homogamy. Looking at hypogamy and hypergamy, the multinomial results (table A.16) emphasize that the odds of marrying down educationally (hypergamy) are higher for employment-related migrants than non-migrants.

Apart from the reason for migration, looking at the spatial characteristics of migrants, migrant stream, and migration type could assist in understanding the movement of migrants and its relationship with EAM. The estimates (figure 2 (B)) indicate that migrants from rural areas (to either urban or rural areas) have a lower likelihood of being educationally matched in the marriage market. Looking at hypogamous and hypergamous marriage, the multinomial results (table A.16, panel c) indicate that migrants from rural areas are more likely to be in educational hypergamy. In contrast, urban-urban migrants have slightly higher odds of hypogamy. Given relatively low levels of education in rural areas, rural-urban migration could result in marrying up educationally.

Lastly, looking for migration type (figure 2 (C), and annexure, table A.16) based on distance, the findings portray that the odds of homogamous (hypergamous) marriage significantly decrease (increases) as migrants move from intra-district to inter-district. Since the IV results are insignificant, EAM might not be significantly impacted based on migration type. Briefly, migrants' characteristics, especially the reason for migration and migration stream, significantly determine the EAM among migrants.

*4.3 Potential Mechanisms*

We attempt to identify what drives a shift towards educational heterogamy and, in some cases, hypogamy among migrants vis-à-vis non-migrants. Following the literature on marriage markets, there are two potential mechanisms. First, the influence of family members and the community (known as a third-party influence) is expected to impact EAM. As, migrant move away from their origin place, their ties with their home community could get weaken. As a result, their marriage preferences could alter or shift towards self-selection.

We cannot directly measure how strong or weak migrant's ties are with their family and community. Thus, we measure how the impact on EAM differs whether the migrant lives in a nuclear or a joint family.[17]. The problem with directly looking at the intersection of migrant and family structure is the possibility of selection bias[18]. In other words, a migrant who choose to be in a nuclear household might already be less traditional thus, have higher odds of being in a heterogamous marriage. As a result, there exist a potential sample selection into nuclear versus joint households for individuals. To overcome the sample selection, we employ the Heckman two-stage model (Heckman, 1979). In the selection model, the dependent variable is type of family structure whereas independent variable are the control variables including at least one exclusion variable[19]. Allendorf (2012) and Niranjan et al., (2005) suggests economic status and age of the household head as determining factors of family structure[20]. Therefore, we include household head's age, share of dependents in the household age, and proportion of expenditure on durables as annual capita consumption expenditure as exclusion variables. We provide argument of exclusion variables as follows. First, household head's age determines the type of family structure as, older (younger) household heads are more (less) likely to be part of joint (nuclear) family structure. Thus, life cycle stage is one of the primary determinants of family structure. Since we control for age-cohorts of individuals, household head's age captures a generational structure of the household which is less likely to affect EAM. Second, share of dependents in the household directly indicates type of family structure as joint family as higher share of dependents. Since number of dependents does not include couple's siblings which could affect their education matching, therefore, share of dependents is less likely to have any causal relationship with assortative mating. Third, proportion of expenditure in durables goods as a share of total annual consumption expenditure determines type of family structure as joint families benefit from economies of scale of shared resources. These expenditure patterns are a result of co-residence and does not determine any earlier educational matching between spouses in the marriage market. Overall, all the three exclusion variables determine family structure either through life cycle stage (household head's age and share of dependents) or economic stage (share of durables expenditure). Meanwhile, the EAM is determined at the time of marriage which is less likely to be affected by exclusion variables considered. The first-stage results are presented in the annexure (table A.18) which indicate positive and significant impact of exclusion variables on likelihood of being in a joint family. In the second stage, we include inverse mills ratio as the covariate.

---

[17] Past studies looking at India's family system (Allendorf, 2012; Mookerjee, 2019) identify the joint family as (i) the presence of the daughter-in-law of the household head or (ii) the mother-in-law when the wife of the household head is present. Using the PLFS (2020-21) data, we identify households as nuclear or joint family structures referring to this identification strategy. Overall, 36 percent of the households in our sample are identified as joint families.
[18] We thank the reviewers for highlighting the issue.
[19] In the Heckman two-stage model, an exclusion variable is the variable that affects the selection into the model (here, type of family structure) but does not impact the main outcome variable (here, EAM).
[20] Other factors that determine the family structure includes, ownership of agricultural land and husband's age (Allendorf, 2012). However, we cannot include husband age as it might also determine our main outcome. Additionally, PLFS does not provide information on ownership of agricultural land.

The results in Table 5 indicate that the likelihood of educational homogamy is lower when the migrant is in a joint family. While the odds are unfavourable in a nuclear family, they are almost 50 per cent higher than in a joint family. Additionally, we interact with migrant status and gender and employ multi-logistic regression to look how pattern differs across gender. We found that within heterogamy, female migrants in a nuclear family have higher odds of being in a hypogamy marriage as compared to homogamy. However, in a joint family, the odds of hypergamy are higher for both migrants and females. Meanwhile, there is no interaction effect of migrant status and gender in a joint family. Therefore, we found that family structure has some influence on EAM. A nuclear family structure increases the odds of being in a non-traditional marital arrangement where the wife has a higher education than the husband. In contrast, a joint family structure pushes toward traditional marital matching. Overall, this mechanism could be one factor which is driving the results for female migrants being in a hypogamous union.

In order to assess whether the results are driven by observations where exclusion variables are outliers, we conduct analyses excluding four borderline cases (presented in table A.21). First, we exclude household where the household head is younger than 30 years. Second, we exclude household where the age of the household head is more than 50 years. Third, to account for outliers, we remove household with extreme durables expenditures (top and bottom 10 percent). Lastly, we remove household with extreme dependency share (top and bottom 10 percent). The sensitivity analyses are presented in the annexure in table A.21. The main findings remain unchanged where migrants in joint family are twice as less likely to be in a homogamous union as migrants in nuclear family.

<center>**<Table 5 here>**</center>

Second, demographic imbalances characterized by sex differential between migrant's origin and destination regions could limit the marriageable pool. Thus, it impacts assortative mating. For instance, post-war, the relative sex ratio changed in favour of men in France, which resulted in caste hypogamy, where men married women from high social castes (Abramitzky et al., 2011). Additionally, the migration of brides from Uttar Pradesh to West Bengal as a result of the low sex ratio in the former state also portrays evidence of how demographic imbalances influence marriage sorting. We attempt to consider the sex-ratio differential for the migrant's origin and destination state. To measure the relative sex ratio[21]First, the migrant's origin and destination states are identified. Then, a state is considered a high (low) sex ratio state if it is more (less) than the mean sex ratio in India based on the latest Census (2011). Lastly, if a migrant's origin state has a low sex ratio and the destination state also has a low sex ratio, then it is labelled as low. Similarly, three other categories are created: low-high, high-low, and high-high, according to sex-ratio states. We interact the migrant status with the migrant's state type (presented in Table 6) using the multi-logit regression. Further, the interaction effect with gender reveals that female migration from a high-high sex ratio state (where the male population is relatively higher than females) increases the likelihood of hypogamy, whereas low-high state migration results in hypergamy. In other words, relatively fewer females in a state (high-sex ratio) could lead to a limited pool of potential brides. Thus, a male educationally matches a female with a higher educational level. This could be because of two factors: (i) males have less bandwidth to be selective towards potential brides, or (ii) females trade the low education of their husbands with some other positive trait such as socioeconomic status or caste. The latter factor has been observed in the case of India by Lin et al. (2020) and Sarkar (2022), which demonstrated that a lower educational level of the husband is compensated by a higher occupation, caste, or socioeconomic status of the husband's or husband's family.

<Table 6 here>

5. Discussion

The study examines the association between migrant status and educational assortative mating (EAM) in India. Using the nationally representative dataset, the Periodic Labour Force Survey (PLFS) (2020-21), we investigate the EAM patterns for India. Overall, there is an increase in educational hypogamy (the wife is more educated than the husband) as we move from the older to the younger cohort. This aligns with previous studies on EAM in India (see, for example, Laha & Pradhan, 2024; Sarkar, 2022). Unlike in developed countries, educational hypogamy is driven by an increase in female educational levels rather than a reversal of gender norms (Lin et al., 2020). Theoretically, industrialization and modernization theory (Blossfled, 2009; Schwartz, 2013) predicts that marriage preference will shift away from homogamy as the economy develops. This could be even more so for migrants for two reasons: their geographical pool of potential spouses widens, and the impact of third-party members in marriage decisions weakens as they move away from their origin place. We found supporting evidence for both these factors empirically.

---

[21] The sex ratio for the prime marriageable cohort (also known as the Marriage Squeeze Index (MSI)) (Kaur et al., 2015) for state*s* is calculated as the ratio of males and females in the group 15 to 34 years multiplied by 100.
$$MSI = \frac{Males_{15-34}}{Females_{15-34}} \times 100$$

Overall, while looking at migration and EAM patterns, we found that educational heterogamy is more prevalent among migrants than among non-migrants. Using logit regression and instrumental variable analysis, the results indicate that female migrants are more likely to engage in educational hypogamy as compared to female non-migrants. Additionally, we look at the heterogeneities in migrant characteristics based on the reason for migration, migration type, migration stream, and couple migrant status. The findings indicate migrants from rural areas and inter-districts have higher odds of educational heterogamy, while urban-urban migrants are likely to be in an educational hypogamy. Since educational hypogamy requires a move away from the traditional marriage norm of homogamy or hypergamy (in the case of traditional societies), urban-urban migrants move towards educational hypogamy.

In a nutshell, results indicate there are some differential features among migrants as compared to non-migrants in terms of EAM. While past studies highlight two potential mechanisms for changing assortative mating, we attempt to explore them empirically. First is third-party influence, which is identified as a type of family structure, joint versus nuclear. In India, the social norms and decision-making power of family, community, and relatives play an important role, especially in marriages. As a result, despite being a migrant, an individual might still be tied to their cultural roots and strictly follow community norms (Chatterjee & Desai, 2020). In our analysis, we found that females in a nuclear family structure are more likely to be in an educationally hypogamous marriage. Thus indicating how family structure determines EAM. Second, geographical pool: as migrants leave their origin place, their potential pool of spouses expands. We attempt to measure this using the sex-ratio differential between the origin and destination state of a migrant. The estimates suggest weak evidence of educational hypogamy as females move from a high-to-high sex ratio state (more males per 100 females). This indicates that a scarcity of potential brides drives males to marry higher-educated females. In past studies on India, educational hypogamy is traded with caste, occupation, or socioeconomic hypergamy. Thus, there is a trade-off between the husband's low educational level and their socioeconomic status. However, whether this exists in the case of migrant status is unknown in India.

The interaction between education and marriage determines a rising level of female education in a country and explains the social distance between different socioeconomic groups in society (Hirschl et al., 2024). The mating patterns among differently educated individuals provide how intergenerational resources are transferred from one generation to another. The educational assortative mating patterns dictate several economic and social outcomes. First, a higher level of educational homogamy among individuals increases disparities among households due to pooled resources among equally wealthy and educated spouses. Thus contributing to rising income and wealth inequality (Boertien & Permanyer, 2019; Schwartz, 2010). A decrease in the likelihood of educational homogamy among migrants could aid in decreasing inequality. Second, a higher level of homogamy is also associated with rigid social boundaries, while heterogamy posits openness of social boundaries (Schwartz, 2013). Therefore, a higher level of heterogamy among migrants than among non-migrants could create a path for decreasing barriers among different education groups and social structures. This is especially relevant for India, where more than 90 and 75 per cent of marriages occur between the same castes and socioeconomic classes (Lin et al., 2020). Third, a rise in hypogamy, where the wife is more educated than the husband, affects intra-household bargaining power. Past studies have found a negative impact on women's autonomy, labour force participation, and domestic violence, among others, as a result of educational hypogamy. This study indicates weak evidence of hypogamy for migrants with some specific characteristics (urban-urban migrants, female migrants, nuclear family, etc.). Thus, further studies can explore whether migrants and non-migrants experience similar outcomes in educational hypogamous unions. However, this backlash effect of hypogamy could be transitory. Given the role of education in career, social, and inter-generational outcomes, a preference towards higher-educated women in the marriage market could rise (Blossfeld & Timm, 2003; Oppenheimer, 1988). However, the male breadwinner notion could still prevail even in dual-earner households as it is deeply rooted in the structure of society.

The study lacks the temporal ordering between migration and marriage as PLFS data does not provide life-course timing of marriage and migration. Therefore, the cross-sectional nature of the data limits the temporal ordering whether migration preceded marriage or vice-versa thus, preventing causal estimates. The central limitation of the study is, our estimates provide association between migrant status and EAM rather than causal estimates. While instrumental variable analysis addresses some identification concerns, but it does not resolve simultaneity or reverse causation. However, by exploring the heterogeneity with respect to gender and reasons of migration provides some clarifications that results are not merely driven by female marriage migrants. For instance, male migrating for employment related reason also exhibits negative association with educational homogamy. Secondly, the potential mechanisms of type of family structure and geographical pool aids in broadening the understanding for how EAM might differ among migrants and non-migrants. Lastly, results based on characteristics of migrants like; migration type and stream indicate differential impact based on distance of migration and type of origin location of migrant. Future studies can establish a life-cycle model using panel data to determine how migrants' preferences differ from that of non-migrants on EAM.

The assortative mating literature for developing countries, particularly India, is relatively recent, which provides ample scope for further research questions. First, it would be interesting to investigate whether female migrants and non-migrants bear different outcomes for women's autonomy in an educationally hypogamous marriage. Second, particularly in dual-earner households, the EAM pattern could vary and impact intra-household bargaining power differently than male-earner households.

# References

Abramitzky, R., Delavande, A., & Vasconcelos, L. (2011). Marrying up: the role of sex ratio in assortative matching. *American Economic Journal: Applied Economics*, *3*(3), 124-157.

Allendorf, K. (2013). Going nuclear? Family structure and young women's health in India, 1992–2006. *Demography*, *50*, 853–880.

Bahl, S., & Sharma, A. (2023). Heterogeneity among migrants, education–occupation mismatch and returns to education. *Regional Studies*, *57*(9), 1851-1865.

Batra, P., & Sharma, A. (2025). International migration and dietary diversity of left-behind households: evidence from India. *Food Security*, 1-15.

Baum, C. F., & Lewbel, A. (2019). Advice on using heteroskedasticity-based identification. *The Stata Journal*, *19*(4), 757-767.

Baum, C., & Schaffer, M. (2024). IVREG2H: Stata module to perform instrumental variables estimation using heteroskedasticity-based instruments.

Beck, A., & González-Sancho, C. (2009). Educational assortative mating and children's school readiness. *Center for Research on Child Wellbeing Working Paper*, *5*.

Becker, G. S. (1993). *A treatise on the family: Enlarged edition*. Harvard University Press.

Bhat, B., & Thakur, S. (2024). Crimes against Women and Migration: Evidence from India. *Available at SSRN 4932507*.

Blossfeld, H. P. (2009). Educational assortative marriage in comparative perspective. *Annual review of sociology*, *35*(1), 513-530.

Blossfeld, H. P., & Timm, A. (Eds.). (2003). *Who marries whom?: Educational systems as marriage markets in modern societies* (Vol. 12). Springer Science & Business Media.

Boertien, D., & Permanyer, I. (2019). Educational assortative mating as a determinant of changing household income inequality: A 21-country study. *European Sociological Review*, *35*(4), 522-537.

Çelikaksoy, A., Nielsen, H. S., & Verner, M. (2006). Marriage migration: just another case of positive assortative matching? *Review of Economics of the Household*, *4*, 253-275.

Chandrasekhar, S., & Sharma, A. (2022). Patterns in Internal Migration and Labour Market Transitions in India. *Centre for the Advanced Study of India*.

Chatterjee, E., & Desai, S. (2021). Physical versus imagined communities: migration and women's autonomy in India. In migration and marriage in Asian contexts (pp. 115–134). Routledge.

Chetty, R., & Hendren, N. (2018). The impacts of neighbourhoods on intergenerational mobility II: County-level estimates. *The Quarterly Journal of Economics*, *133*(3), 1163–1228.

Choi, K. H., & Mare, R. D. (2012). International migration and educational assortative mating in Mexico and the United States. *Demography*, *49*(2), 449–476.

Dang, H. A. H., & Rogers, F. H. (2016). The decision to invest in child quality over quantity: Household size and household investment in education in Vietnam. *The World Bank Economic Review*, *30*(1), 104–142.

Davis, K. (1941). Intermarriage in caste societies. *American Anthropologist*, *43*(3), 376–395.

De Haas, H. (2010). The internal dynamics of migration processes: A theoretical inquiry. *Journal of ethnic and migration studies*, *36*(10), 1587–1617.

Dziadula, E., & Zavodny, M. (2024). Finding love abroad: who marries a migrant and what do they gain?. *Review of Economics of the Household*, *22*(4), 1371-1396.

Eggert, W., Krieger, T., & Meier, V. (2010). Education, unemployment and migration. *Journal of Public Economics*, *94*(5-6), 354-362.

Emran, M. S., & Hou, Z. (2013). Access to markets and rural poverty: evidence from household consumption in China. Review of Economics and Statistics, 95(2), 682-697.

Esteve, A., & Bueno, X. (2012). Marrying after migration: Assortative mating among Moroccans in Spain. *Genus*, *68*(1), 41-63.

Esteve, A., García-Román, J., & Permanyer, I. (2012). The gender-gap reversal in education and its effect on union formation: the end of hypergamy? *Population and Development Review*, *38*(3), 535-546.

Fan, C. C., & Li, L. (2002). Marriage and migration in transitional China: a field study of Gaozhou, western Guangdong. *Environment and Planning A*, *34*(4), 619-638.

Goel, P. A., & Barua, R. (2023). Female education, marital assortative mating, and dowry: Theory and evidence from districts of India. *Journal of Demographic Economics*, *89*(2), 183-209.

Greenwood, M. J. (1997). Internal migration in developed countries. *Handbook of population and family economics*, *1*, 647–720.

Hirschl, N., Schwartz, C. R., & Boschetti, E. (2024). Eight decades of educational assortative mating: A research note. *Demography*, *61*(5), 1293-1307.

Hu, Q. (2024). Social status and marriage markets: Evaluating a Hukou policy in China. *Review of Economics of the Household*, *22*(2), 477-509.

Jia, N., Molloy, R., Smith, C., & Wozniak, A. (2023). The economics of internal migration: Advances and policy questions. *Journal of Economic Literature*, *61*(1), 144-180.

Kain, J. F. (1968). Housing segregation, negro employment, and metropolitan decentralization. *The quarterly journal of economics*, *82*(2), 175–197.

Kalmijn, M. (1991). Status homogamy in the United States. *American Journal of Sociology*, *97*(2), 496–523.

Kalmijn, M. (1998). Intermarriage and homogamy: Causes, patterns, trends. *Annual review of sociology*, *24*(1), 395–421.

Kaur, R. (2012). Marriage and migration: Citizenship and marital experience in cross-border marriages between Uttar Pradesh, West Bengal and Bangladesh. *Economic and Political Weekly*, 78–89.

Kaur, R., Bhalla, S. S., Agarwal, M. K., & Ramakrishnan, P. (2015). Sex Ratio Imbalances and Marriage Squeeze in India: 2000-2050.

Laha, P., & Pradhan, M. R. (2024). Patterns and Predictors of Educational Hypogamy in India. *Marriage & Family Review*, *60*(8), 615-631.

Lewbel, A. (2012). Using heteroscedasticity to identify and estimate mismeasured and endogenous regressor models. *Journal of business & economic statistics*, *30*(1), 67–80.

Lewbel, A. (2018). Identification and estimation using heteroscedasticity without instruments: The binary endogenous regressor case. *Economics Letters*, *165*, 10-12.

Lin, Z., Desai, S., & Chen, F. (2020). The emergence of educational hypogamy in India. *Demography*, *57*(4), 1215-1240.

Lucas, R. E. (2016). Internal migration in developing economies: an overview of recent evidence. *Geopolitics, History, and International Relations*, *8*(2), 159–191.

Mahesh, M. (2020). The effect of remittances on crime in India. *IZA Journal of Labor Policy*, *10*(1).

Mookerjee, S. (2019). Gender-neutral inheritance laws, family structure, and women's status in India. *The World Bank Economic Review*, *33*(2), 498–515.

Mu, Z., & Yeung, W. J. J. (2021). Internal migration, marriage timing and assortative mating: A mixed-method study in China. In *Migration and Marriage in Asian Contexts* (pp. 52–74). Routledge.

Niranjan, S., Nair, S., & Roy, T. K. (2005). A socio-demographic analysis of the size and structure of the family in India. *Journal of Comparative Family Studies*, *36*(4), 623-652.

NSO. (2022). *Annual Report, Periodic Labour Force Survey (2020–21)*. Government of India. Ministry of Statistics and Programme Implementation.

Oppenheimer, V. K. (1988). A theory of marriage timing. *American journal of sociology*, *94*(3), 563–591.

Potarca, G., & Bernardi, L. (2017, April). Educational Hypogamy and Intermarriage: Are Immigrants Drivers of Change in Educational Sorting? In *PAA 2017 Annual Meeting*. PAA.

Preston, V., & McLafferty, S. (1999). Spatial mismatch research in the 1990s: progress and potential. *Papers in regional science*, *78*, 387-402.

Qian, Z., & Lichter, D. T. (2007). Social boundaries and marital assimilation: Interpreting trends in racial and ethnic intermarriage. *American Sociological Review*, *72*(1), 68-94.

Sarkar, K. (2022). Can status exchanges explain educational hypogamy in India? *Demographic Research*, *46*, 809–848.

Sarkar, S. (2024). Local crime and early marriage: Evidence from India. *The Journal of Development Studies*, *60*(5), 763–787.

Schwartz, C. R. (2010). Earnings inequality and the changing association between spouses' earnings. *American journal of sociology*, *115*(5), 1524-1557.

Schwartz, C. R. (2013). Trends and variation in assortative mating: Causes and consequences. *Annual Review of Sociology*, *39*(1), 451-470.

Sharma, A., & Das, M. (2018). Migrant networks in the urban labour market: Evidence from India. *The Journal of Development Studies*, *54*(9), 1593–1611.

South, S. J., & Crowder, K. D. (1999). Neighbourhood effects on family formation: Concentrated poverty and beyond. *American Sociological Review*, *64*(1), 113–132.

Suedekum, J., Wolf, K., & Blien, U. (2014). Cultural diversity and local labour markets. *Regional Studies*, *48*(1), 173-191.

Roychowdhury, P., & Dhamija, G. (2024). Educational hypogamy and female employment in rural India. *Empirical Economics*, 1–39.

Roychowdhury, P., & Dhamija, G. (2022). Don't cross the line: Bounding the causal effect of hypergamy violation on domestic violence in India. *Journal of the Royal Statistical Society Series A: Statistics in Society*, *185*(4), 1952-1978.

Smits, J., Ultee, W., & Lammers, J. (1998). Educational homogamy in 65 countries: An explanation of differences in openness using country-level explanatory variables. *American Sociological Review*, 264-285.

Sun, F., Xiaohui, L., & Xiao, J. J. (2021). Who married whom? Rural-urban migration and mate selection in China. *Family Science Review*, *25*(1), 36-38.

UNFPA India. (2022, July 25). *Child Marriage in India: Key Insights from the NFHS-5 (2019-21)* (Analytical Paper Series 1). Retrieved from https://india.unfpa.org/en/publications/analytical-paper-series-1-child-marriage-india-key-insights-nfhs-5-2019-21

Thai, H. C. (2008). *For better or for worse: Vietnamese international marriages in the new global economy*. Rutgers University Press.

Zhang, L., & Tan, X. (2021). Educational assortative mating and health: a study in Chinese internal migrants. *International Journal of Environmental Research and Public Health*, *18*(4), 1375.

## 6. Migration and Educational Assortative Mating in India: How Geographic Mobility Shapes Marriage Markets

**Figures**

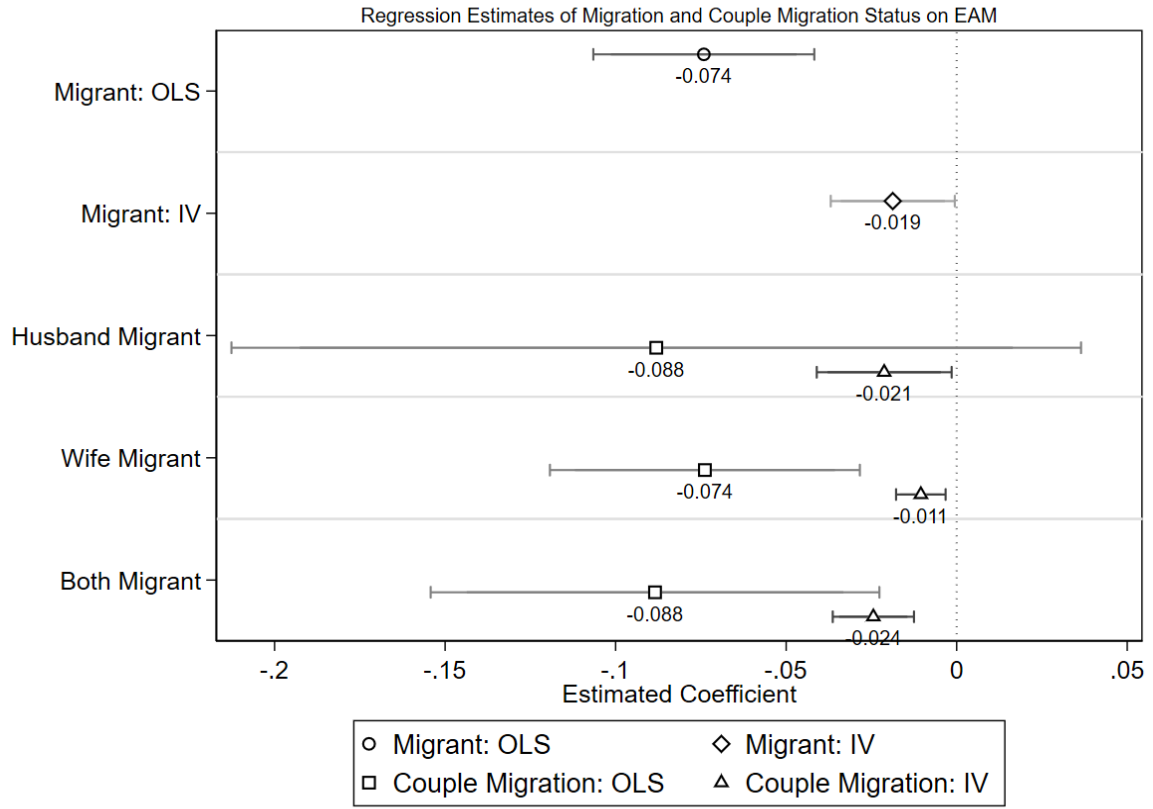

Figure 1: Regression Estimates of Migration and Couple Migration Status on EAM

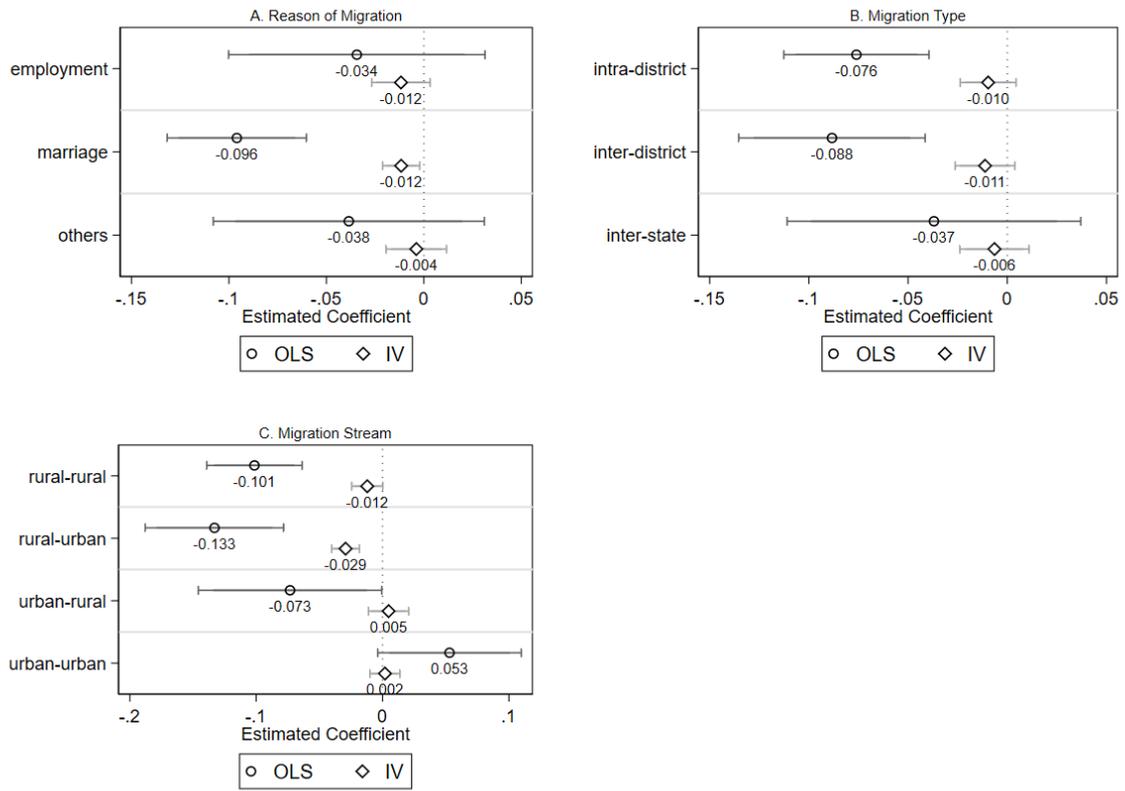

Figure 2: Regression Estimates of Reason of Migration, Migration Type, and Migration Stream on EAM

7. **Migration and Educational Assortative Mating in India: How Geographic Mobility Shapes Marriage Markets**

8. **Tables**

**Table 1: Distribution of Educational Assortative Mating and Age Cohorts**

| Type of Marriage/ Age-cohorts | | 15-24 | 25-34 | 35-44 | 45-59 | Total |
|---|---|---|---|---|---|---|
| | | **Panel (a) Overall** | | | | |
| Homogamy | | 35.53 | 36.64 | 38.55 | 43.54 | 39.47 |
| Heterogamy | | 64.47 | 63.37 | 61.45 | 56.46 | 60.53 |
| | Hypogamy | 27.03 | 25.03 | 18.7 | 12.59 | 19.02 |
| | Hypergamy | 37.44 | 38.34 | 42.75 | 43.87 | 41.51 |
| | | **Panel (b) Non-migrant** | | | | |
| Homogamy | | 37.81 | 36.22 | 38.43 | 43.63 | 39.69 |
| Heterogamy | | 62.19 | 63.79 | 61.57 | 56.37 | 60.32 |
| | Hypogamy | 25.85 | 25.27 | 19.78 | 12.97 | 19.11 |
| | Hypergamy | 36.34 | 38.52 | 41.79 | 43.4 | 41.21 |
| | | **Panel (c) Migrant** | | | | |
| Homogamy | | 33.96 | 37.1 | 38.7 | 43.41 | 39.2 |
| Heterogamy | | 66.04 | 62.91 | 61.3 | 56.59 | 60.8 |
| | Hypogamy | 27.84 | 24.77 | 17.34 | 12.04 | 18.92 |
| | Hypergamy | 38.2 | 38.14 | 43.96 | 44.55 | 41.88 |
| Total | | 100 | 100 | 100 | 100 | 100 |

Source: Authors' calculations using PLFS (2020-21)

Note: (i) Frequency weights are employed.

(ii) Homogamy (heterogamy) refers to an equal (unequal) level of education between wife and husband. Hypogamy (hypergamy) refers to the wife's (husband) level of education exceeding the husband's (wife's) level of education.

**Table 2: Distribution of Educational Assortative Mating and Couple Migration Status**

| Type of Marriage/ Couple Migration Status | Both non-migrant | Husband-migrant | Wife-migrant | Both Migrant | Total |
|---|---|---|---|---|---|
| Homogamy | 41.29 | 38.39 | 39.03 | 38.38 | 39.47 |
| Heterogamy | 58.71 | 61.61 | 60.97 | 61.62 | 60.53 |
|     Hypogamy | 19.26 | 26.4 | 18.39 | 21.02 | 19.02 |
|     Hypergamy | 39.45 | 35.21 | 42.58 | 40.6 | 41.51 |
| Total | 100 | 100 | 100 | 100 | 100 |

Source: Authors' calculations using PLFS (2020-21)

Note: (i) Frequency weights are employed.

(ii) Homogamy (heterogamy) refers to an equal (unequal) level of education between wife and husband. Hypogamy (hypergamy) refers to the wife's (husband) level of education exceeding the husband's (wife's) level of education.

## Table 3: Regression Estimates for Educational Homogamy among Migrants

| Dependent Variable: Educational Homogamy | (1) Logit | (2) Logit | (3) Logit | (4) IV | (5) IV | (6) IV |
|---|---|---|---|---|---|---|
| | | Rural | Urban | | Rural | Urban |
| **Panel (a): Migrant Status** | | | | | | |
| Migrant | -0.074*** | -0.066*** | -0.054** | -0.019** | -0.009 | -0.023* |
| | (0.017) | (0.023) | (0.024) | (0.009) | (0.010) | (0.013) |
| Observations | 128,400 | 77,177 | 51,223 | 125,996 | 75,863 | 50,133 |
| Marginal effect of Migrant | -0.016 | -0.013 | -0.012 | -0.019 | -0.009 | -0.024 |
| Mean Dep Variable | 0.356 | 0.348 | 0.369 | 0.356 | 0.348 | 0.369 |
| F-statistic | - | - | - | 216.92 | 108.51 | 93.20 |
| Under-identification Test | - | - | - | 6382.303 | 4206.77 | 5844.58 |
| **Panel (b): Couple Migrant Status** | | | | | | |
| Base cat.: Non-migrant | | | | | | |
| Husband Migrant | -0.088 | -0.012 | -0.137* | -0.021** | -0.004 | -0.037*** |
| | (0.064) | (0.096) | (0.083) | (0.010) | (0.014) | (0.014) |
| Wife Migrant | -0.074*** | -0.029 | -0.107*** | -0.010*** | -0.005 | -0.020*** |
| | (0.023) | (0.031) | (0.035) | (0.004) | (0.005) | (0.006) |
| Both Migrant | -0.088*** | -0.078 | -0.087** | -0.024*** | -0.017* | -0.021*** |
| | (0.034) | (0.059) | (0.042) | (0.006) | (0.009) | (0.007) |
| Observations | 128,400 | 77,177 | 51,223 | 125,996 | 75,863 | 50,133 |
| Marginal effect of (a) Husband Migrant | -0.019 | -0.002 | -0.031 | -0.021 | -0.004 | -0.037 |
| (b) Wife Migrant | -0.016 | -0.006 | -0.024 | -0.011 | -0.005 | -0.020 |
| (c) Both Migrant | -0.019 | -0.016 | -0.020 | -0.024 | -0.017 | -0.021 |
| Mean Dep Variable | 0.356 | 0.348 | 0.369 | 0.356 | 0.348 | 0.369 |
| F-statistic | - | - | - | 879.211 | 2257.920 | 989.524 |
| Under-identification Test | - | - | - | 2.8e+04 | 6441.506 | 2.2e+04 |

Source: Authors' calculation based on PLFS Survey, 2020-21

Note: (i) *** signals significant at 1% level, ** signals significant at 5% level, and * signals significant at 10% level.

(ii) Standard errors are clustered at FSU levels for logit model.

(iii) The analysis controls for education categories, age-cohorts, gender, sector, social group, religion, and eligible men-to-women ratio in rural and urban areas, 36 states and union territories.

(iv) The Lewbel IV models contain three instrumental variables, namely, migrant network, crime, and unemployment rate in rural and urban areas. The first stage results are reported in the annexure (table A.10 for migrant status and table A.11 for couple migration status).

**Table 4: Multinomial Regression Estimates for Educational Homogamy among Migrants**

| Base Category: Homogamous Union | (1) Hypogamous | (2) Hypergamous |
|---|---|---|
| **Panel (a): Migrant Status** | | |
| Migrant | 0.030 | 0.089*** |
|  | (0.021) | (0.018) |
| Observations | 128,400 | 128,400 |
| **Panel (b): Interaction with Gender** | | |
| Migrant | -0.017 | 0.079*** |
|  | (0.034) | (0.029) |
| Migrant*Female | 0.073* | 0.016 |
|  | (0.039) | (0.033) |
| Observations | 128,400 | 128,400 |
| **Panel (c): Couple Migrant Status** | | |
| Base cat: Both non-migrant | | |
| Husband Migrant | 0.125 | 0.057 |
|  | (0.078) | (0.073) |
| Wife Migrant | 0.069** | 0.073*** |
|  | (0.031) | (0.026) |
| Both Migrant | -0.025 | 0.143*** |
|  | (0.044) | (0.037) |
| Observations | 128,400 | 128,400 |

Source: Authors' calculation based on PLFS Survey, 2020-21

Note: (i) *** signals significant at 1% level, ** signals significant at 5% level, and * signals significant at 10% level.

(ii) Standard errors are clustered at FSU levels.

(iii) The analysis controls for education categories, age-cohorts, gender, sector, social group, religion, and the ratio of eligible men to women in rural and urban areas, 36 states and union territories.

**Table 5: Mechanism: Regression Estimates for Educational Homogamy: Based on Family Structure (Nuclear and Joint Family)**

|  | (1) | (2) | (3) | (4) | (5) | (6) |
|---|---|---|---|---|---|---|
|  | Nuclear | Joint | Nuclear | | Joint | |
| Dependent Variable: Homogamous Union |  |  | Hypogamy | Hypergamy | Hypogamy | Hypergamy |
| Base cat.: Non-migrant |  |  |  |  |  |  |
| Migrant | -0.051** | -0.094*** | -0.063 | 0.014 | 0.057 | 0.145*** |
|  | (0.021) | (0.027) | (0.042) | (0.034) | (0.061) | (0.054) |
| Female | -0.061*** | -0.040** | -0.078*** | 0.075*** | -0.051 | 0.097*** |
|  | (0.014) | (0.019) | (0.025) | (0.020) | (0.031) | (0.027) |
| Migrant*Female | - | - | 0.108** | 0.078** | 0.015 | -0.062 |
|  |  |  | (0.047) | (0.039) | (0.071) | (0.062) |
| Observations | 80,886 | 47,400 | 80,886 | 80,886 | 47,400 | 47,400 |

Source: Authors' calculation based on PLFS Survey, 2020-21

Note: (i) *** signals significant at 1% level, ** signals significant at 5% level, and * signals significant at 10% level.

(ii) Standard errors are clustered at FSU levels.

(iii) The analysis controls for education categories, age-cohorts, gender, sector, social group, religion, and the ratio of eligible men to women in rural and urban areas, 36 states and union territories. The inverse mills ratio are also included.

(iv) Columns (3), (4), (5), and (6) consist of multinominal regression where the base category of the dependent variable is educational homogamy.

**Table 6: Mechanism: Multinominal Regression Estimates for Educational Homogamy: Based on Sex Differential between Origin and Destination State**

| Base Category: Homogamous Union | (1) Hypogamy | (2) Hypergamy |
|---|---|---|
| **Base cat.: Non-migrant Male** | | |
| Low-Low*Female | 0.057 | 0.003 |
| | (0.044) | (0.037) |
| Low-High*Female | 0.103 | 0.196** |
| | (0.103) | (0.084) |
| High-Low*Female | 0.051 | -0.054 |
| | (0.172) | (0.128) |
| High-High*Female | 0.154* | 0.026 |
| | (0.085) | (0.070) |
| Female | -0.051*** | 0.113*** |
| | (0.019) | (0.016) |
| **Base cat.: Non-migrant** | | |
| Low-Low | -0.011 | 0.090*** |
| | (0.040) | (0.034) |
| Low-High | 0.072 | 0.181 |
| | (0.136) | (0.120) |
| High-Low | -0.076 | -0.032 |
| | (0.119) | (0.088) |
| High-High | -0.049 | 0.054 |
| | (0.091) | (0.073) |
| Observations | 128,399 | 128,399 |

Source: Authors' calculation based on PLFS Survey, 2020-21

Note: (i) *** signals significant at 1% level, ** signals significant at 5% level, and * signals significant at 10% level.

(ii) Standard errors are clustered at FSU levels.

(iii) The analysis controls for education categories, age-cohorts, gender, sector, social group, religion, and the ratio of eligible men to women in rural and urban areas, 36 states and union territories.

9. **Migration and Educational Assortative Mating in India: How Geographic Mobility Shapes Marriage Markets**
10. **Annexure**
11. **A.1 Couple Identification Strategy**

In order to identify the educational assortative matching, we first identify the couples in the data. The Periodic Labour Force Survey (PLFS) does not provide couple identifiers; therefore, we attempt to identify couples using the "relation to head" to household and "marital status" of individuals. To do so, we restrict our sample to married individuals whose relation to the head is one of the following: spouse of the head, married child, and spouse of the married child. We also generate a couple serial number (CSN) for our reference to generate a unique couple identifier ID.

First, we identify the household head couple by restricting our data to the household head and spouse of the household head. We call them the "first generation couple" and assign them CSN (=1). We can identify 77,163 couples (154,326 individuals) and create spouse education as a variable for reference.

Second, to identify a "second generation couple", a married child and their spouse. For simplicity, we first look for households with only one second-generation couple, assign them CSN (=2), and create spouse education. Considering only the households with only one second-generation couple present, we can identify 14,045 second-generation couples (28,190 individuals).

Next, we identify the households with more than one second-generation couple. The caveat is that there are households where a married child exceeds the spouse of a married child or vice versa, which indicates these individuals are married, but their spouse is not present in the household. Therefore, we drop individuals whose spouses are absent from the households and refer to them as "loners". We drop 1,237 individuals. Using the person serial number and relationship to head, we identified 3,518 (7,036 individuals) second-generation couples with up to five married children and assigned them 3 to 7 CSNs. Lastly, we assign a couple identifier ID, which is the concatenate by joining household is (HHID), person serial number (PSN), and CSN. The sample is detailed in Table A.1.

## 12. A.2 Annexure Tables

### Table A.1: Details about the sample

| Sample Head | Count |
|---|---|
| Total Married Individuals | 202,861 |
| Identified Individuals (Couples) | 189,552 (94,776) |
| Dropping individuals in same-sex unions = 74 | |
| Identified Individuals (Couples) | 189, 478 (94, 739) |
| Dropping covid migrants, guest members, and international migrants | |
| Final Sample Individuals (Couples) | 185,284 (92, 642) |

Source: Authors' calculations based on PLFS (2020-21)

Table A.2: Distribution of Educational Assortative Mating and Age Cohorts Across Sectors

| Type of Marriage/ Age-cohorts | | 15-24 | 25-34 | 35-44 | 45-59 | Total |
|---|---|---|---|---|---|---|
| **Panel (a) Rural** | | | | | | |
| Homogamy | | 34.77 | 35.66 | 38.31 | 45.57 | 39.65 |
| Heterogamy | | 65.23 | 64.33 | 61.69 | 54.43 | 60.36 |
| | Hypogamy | 26.38 | 23.57 | 16.56 | 10.2 | 17.26 |
| | Hypergamy | 38.85 | 40.76 | 45.13 | 44.23 | 43.1 |
| **Panel (b) Urban** | | | | | | |
| Homogamy | | 38.52 | 39.11 | 39.12 | 38.96 | 39.03 |
| Heterogamy | | 61.48 | 60.9 | 60.89 | 61.04 | 60.97 |
| | Hypogamy | 29.57 | 28.71 | 23.76 | 17.97 | 23.34 |
| | Hypergamy | 31.91 | 32.19 | 37.13 | 43.07 | 37.63 |

Source: Authors' calculations using PLFS (2020-21)

Notes: (i) Frequency weights are employed.
(ii) Homogamy (heterogamy) refers to an equal (unequal) level of education between wife and husband. Hypogamy (hypergamy) refers to the wife's (husband) level of education exceeding the husband's (wife's) level of education.

# Table A.3: Summary Statistics (Working-age population)

| Variable | Observation | Mean | Observation | Mean | Observation | Mean |
|---|---|---|---|---|---|---|
| | Overall | | Non-migrants | | Migrants | |
| **Educational Assortative Matching** | | | | | | |
| Homogamy | 160476 | 0.358 | 92010 | 0.36 | 68466 | 0.355 |
| Hypogamy | 160476 | 0.206 | 92010 | 0.205 | 68466 | 0.208 |
| Hypergamy | 160476 | 0.436 | 92010 | 0.435 | 68466 | 0.437 |
| **Education Categories** | | | | | | |
| Illiterate | 161973 | 0.195 | 92812 | 0.159 | 69161 | 0.244 |
| Primary | 161973 | 0.184 | 92812 | 0.183 | 69161 | 0.186 |
| Middle | 161973 | 0.238 | 92812 | 0.255 | 69161 | 0.215 |
| Secondary | 161973 | 0.141 | 92812 | 0.152 | 69161 | 0.127 |
| Higher Secondary | 161973 | 0.115 | 92812 | 0.122 | 69161 | 0.106 |
| Graduate & Above | 161973 | 0.126 | 92812 | 0.128 | 69161 | 0.123 |
| **Age-cohorts** | | | | | | |
| 15-24 | 161973 | 0.065 | 92812 | 0.05 | 69161 | 0.087 |
| 25-34 | 161973 | 0.26 | 92812 | 0.246 | 69161 | 0.28 |
| 35-44 | 161973 | 0.322 | 92812 | 0.324 | 69161 | 0.32 |
| 45-59 | 161973 | 0.352 | 92812 | 0.38 | 69161 | 0.314 |
| **Gender** | | | | | | |
| Male | 161973 | 0.476 | 92812 | 0.719 | 69161 | 0.15 |
| Female | 161973 | 0.524 | 92812 | 0.281 | 69161 | 0.85 |
| **Sector** | | | | | | |
| Rural | 161973 | 0.58 | 92812 | 0.597 | 69161 | 0.558 |
| Urban | 161973 | 0.42 | 92812 | 0.403 | 69161 | 0.442 |
| **Religion** | | | | | | |
| Hindu | 161973 | 0.766 | 92812 | 0.729 | 69161 | 0.817 |
| Islam | 161973 | 0.128 | 92812 | 0.137 | 69161 | 0.116 |
| Others | 161973 | 0.106 | 92812 | 0.135 | 69161 | 0.067 |
| **Social Group** | | | | | | |
| Scheduled Tribe (ST) | 161973 | 0.139 | 92812 | 0.167 | 69161 | 0.102 |
| Scheduled Caste (SC) | 161973 | 0.182 | 92812 | 0.176 | 69161 | 0.19 |
| Other Backward Classes | 161973 | 0.398 | 92812 | 0.387 | 69161 | 0.413 |
| Other | 161973 | 0.281 | 92812 | 0.27 | 69161 | 0.295 |
| MPCE | 161973 | 8.849 | 92812 | 8.816 | 69161 | 8.892 |
| Eligible Marriage Pool (Rural) | 129710 | 0.973 | 74796 | 0.971 | 54914 | 0.976 |
| Eligible Marriage Pool (Urban) | 131878 | 0.978 | 75764 | 0.976 | 56114 | 0.982 |

Source: Authors' calculations based on PLFS (2020-21)

Note: The summary statistics are calculated for matched individuals (only couples).

Homogamy refers to an equal level of education between wife and husband. Hypogamy (hypergamy) refers to the wife's (husband) level of education being more than the husband's (wife's) level of education.

Table A.4: Percentage distribution of wives' and husbands' education levels Across Migrant Status

Panel (a): Overall

| Husband's education | Wife's education | | | | | | | | Total |
|---|---|---|---|---|---|---|---|---|---|
| | Not literate | Below Primary | Primary | Middle | Secondary | Higher Secondary | Undergraduate | Post-graduate | |
| Not literate | 15.17 | 0.8 | 1.21 | 0.94 | 0.3 | 0.08 | 0.03 | 0 | 18.53 |
| Below Primary | 2.46 | 1.94 | 0.82 | 0.82 | 0.27 | 0.09 | 0.03 | 0 | 6.43 |
| Primary | 4.46 | 1.41 | 3.78 | 2.51 | 0.95 | 0.39 | 0.11 | 0.01 | 13.62 |
| Middle | 5.15 | 1.56 | 4.4 | 7.78 | 2.32 | 1.43 | 0.52 | 0.07 | 23.22 |
| Secondary | 1.96 | 0.6 | 1.85 | 3.35 | 3.26 | 1.63 | 0.75 | 0.16 | 13.58 |
| Higher Secondary | 0.77 | 0.26 | 0.84 | 2.55 | 2.39 | 2.82 | 1.45 | 0.36 | 11.44 |
| Undergraduate | 0.34 | 0.1 | 0.37 | 1.28 | 1.17 | 2.11 | 3.58 | 0.97 | 9.91 |
| Post-graduate | 0.06 | 0.01 | 0.06 | 0.23 | 0.27 | 0.44 | 1.07 | 1.14 | 3.27 |
| Total | 30.37 | 6.69 | 13.32 | 19.46 | 10.92 | 8.99 | 7.53 | 2.71 | 100 |

Panel (b): Non-migrant

| Husband's education | Wife's education | | | | | | | | Total |
|---|---|---|---|---|---|---|---|---|---|
| | Not literate | Below Primary | Primary | Middle | Secondary | Higher Secondary | Undergraduate | Post-graduate | |
| Not literate | 15.43 | 0.82 | 1.21 | 0.95 | 0.31 | 0.08 | 0.03 | 0 | 18.82 |
| Below Primary | 2.48 | 2.08 | 0.88 | 0.89 | 0.31 | 0.09 | 0.03 | 0 | 6.75 |
| Primary | 4.27 | 1.4 | 3.78 | 2.57 | 0.97 | 0.4 | 0.13 | 0.01 | 13.53 |
| Middle | 5.05 | 1.65 | 4.34 | 8 | 2.35 | 1.47 | 0.5 | 0.07 | 23.44 |
| Secondary | 1.98 | 0.63 | 1.87 | 3.48 | 3.33 | 1.6 | 0.69 | 0.17 | 13.75 |

| Husband's education | | | | | | | | | |
|---|---|---|---|---|---|---|---|---|---|
| Higher Secondary | 0.72 | 0.27 | 0.82 | 2.63 | 2.42 | 2.73 | 1.41 | 0.32 | 11.32 |
| Undergraduate | 0.3 | 0.1 | 0.36 | 1.33 | 1.21 | 2.01 | 3.35 | 0.84 | 9.5 |
| Post-graduate | 0.05 | 0.01 | 0.05 | 0.22 | 0.25 | 0.41 | 0.89 | 1 | 2.89 |
| Total | 30.29 | 6.96 | 13.3 | 20.05 | 11.14 | 8.79 | 7.04 | 2.42 | 100 |

| Panel (c): Migrant | | | | | | | | | |
|---|---|---|---|---|---|---|---|---|---|
| Husband's education | Wife's education | | | | | | | | |
| | Not literate | Below Primary | Primary | Middle | Secondary | Higher Secondary | Undergraduate | Post-graduate | Total |
| Not literate | 14.87 | 0.78 | 1.2 | 0.93 | 0.29 | 0.08 | 0.03 | 0 | 18.18 |
| Below Primary | 2.43 | 1.77 | 0.75 | 0.74 | 0.22 | 0.1 | 0.02 | 0 | 6.04 |
| Primary | 4.68 | 1.43 | 3.79 | 2.43 | 0.93 | 0.37 | 0.09 | 0.01 | 13.73 |
| Middle | 5.26 | 1.46 | 4.46 | 7.51 | 2.29 | 1.38 | 0.54 | 0.07 | 22.96 |
| Secondary | 1.93 | 0.57 | 1.83 | 3.2 | 3.18 | 1.67 | 0.83 | 0.15 | 13.37 |
| Higher Secondary | 0.84 | 0.24 | 0.86 | 2.45 | 2.35 | 2.93 | 1.49 | 0.42 | 11.59 |
| Undergraduate | 0.38 | 0.09 | 0.38 | 1.23 | 1.12 | 2.23 | 3.85 | 1.11 | 10.4 |
| Post-graduate | 0.07 | 0.01 | 0.07 | 0.24 | 0.28 | 0.47 | 1.28 | 1.3 | 3.73 |
| Total | 30.46 | 6.36 | 13.35 | 18.74 | 10.67 | 9.23 | 8.13 | 3.07 | 100 |

Source: Authors' calculations using PLFS (2020-21)
Note: Frequency weights are employed.

**Table A.5: Exclusion Criteria: Regression Estimates of Impact of Instrumental Variables on Education Assortative Matching (only for non-migrants)**

| Dependent Variable: Homogamous Union | (1) |
|---|---|
| Crime | -0.019 |
|  | (0.024) |
| Unemployment rate (Rural) | 0.003 |
|  | (0.009) |
| Unemployment rate (Urban) | 0.009 |
|  | (0.007) |
| Migrant Network | 0.030 |
|  | (0.034) |
| Observations | 72,791 |

Source: Authors' calculation based on PLFS Survey, 2020-21

Note: (i) *** signals significant at 1% level, ** signals significant at 5% level, and * signals significant at 10% level.
(ii) The analysis controls for education categories, age-cohorts, gender, sector, social group, religion, and eligible men-to-women ratio in rural and urban areas, 36 states and union territories.
(iii) Standard errors are clustered at the FSU level.

**Table A.6: Sensitivity Check: Regression Estimates for Educational Homogamy among Migrants (excluding marriage migrants)**

| Dependent Variable: Homogamous Union | (1) | (2) |
|---|---|---|
|  | OLS | IV |
| Migrant | -0.030 | -0.018*** |
|  | (0.030) | (0.007) |
| Observations | 85,608 | 84,032 |

Source: Authors' calculation based on PLFS Survey, 2020-21

Note: (i) *** signals significant at 1% level, ** signals significant at 5% level, and * signals significant at 10% level.
(ii) The analysis controls for education categories, age-cohorts, gender, sector, social group, religion, and eligible men-to-women ratio in rural and urban areas, 36 states and union territories.
(iii) Standard errors are clustered at the FSU level.

**Table A.7: Sensitivity Check: Regression Estimates for Educational Homogamy among Migrants (Age > 23 and Age >30)**

| Dependent Variable: Educational Homogamy | (1) Logit | (2) Logit Rural | (3) Logit Urban |
|---|---|---|---|
| **Panel (a): Age >23** | | | |
| Migrant | -0.068*** | -0.059** | -0.052** |
|  | (0.017) | (0.025) | (0.024) |
| Observations | 121,741 | 72,447 | 49,294 |
| **Panel (b): Age> 30** | | | |
| Migration | -0.074*** | -0.065** | -0.057** |
|  | (0.019) | (0.028) | (0.027) |
| Observations | 98,341 | 57,731 | 40,610 |

Source: Authors' calculation based on PLFS Survey, 2020-21

Note: (i) *** signals significant at 1% level, ** signals significant at 5% level, and * signals significant at 10% level.

(ii) Standard errors are clustered at FSU levels.

(iii) The analysis controls for education categories, age-cohorts, gender, sector, social group, religion, and eligible men-to-women ratio in rural and urban areas, 36 states and union territories.

**Table A.8: Heterogeneity Analysis: Regression Estimates for Educational Homogamy: Based on Reason of Migration and Migration Type**

| Dependent Variable: Homogamous Union | (1) Logit | (2) Logit Rural | (3) Logit Urban | (4) IV | (5) IV Rural | (6) IV Urban |
|---|---|---|---|---|---|---|
| **Panel (a): Reason for Migration** | | | | | | |
| Base cat.: Non-migrant | | | | | | |
| Employment | -0.034 | -0.081 | 0.003 | -0.012 | -0.016 | -0.002 |
|  | (0.034) | (0.066) | (0.039) | (0.008) | (0.013) | (0.009) |
| Marriage | -0.096*** | -0.069*** | -0.100*** | 0.012** | -0.005 | 0.021*** |
|  | (0.018) | (0.025) | (0.027) | (0.005) | (0.006) | (0.007) |
| Other | -0.038 | -0.025 | -0.029 | -0.004 | -0.002 | -0.008 |
|  | (0.035) | (0.064) | (0.043) | (0.008) | (0.012) | (0.010) |
| Observations | 128,400 | 77,177 | 51,223 | 125,996 | 75,863 | 50,133 |
| F-statistic |  |  |  | 858.463 | 730.30 | 420.774 |
| Under-identification Test |  |  |  | 3.4e+04 | 2.0e+04 | 2.0e+04 |
| **Panel (b): Migration Type** | | | | | | |
| Base cat.: Non-migrant | | | | | | |
| Intra-district | -0.076*** | -0.055** | -0.065** | -0.010 | -0.001 | -0.015 |
|  | (0.019) | (0.026) | (0.028) | (0.007) | (0.008) | (0.012) |
| Inter-district | -0.088*** | -0.073** | -0.077** | -0.011 | 0.001 | -0.016 |
|  | (0.024) | (0.035) | (0.033) | (0.008) | (0.009) | (0.011) |
| Inter-state | -0.037 | -0.117** | 0.016 | -0.007 | -0.017 | 0.003 |
|  | (0.038) | (0.058) | (0.048) | (0.009) | (0.011) | (0.012) |
| Observations | 128,400 | 77,177 | 51,223 | 125,996 | 75,863 | 50,133 |
| F-statistic |  |  |  | 239.39 | 249.01 | 95.27 |
| Under-identification Test |  |  |  | 1.5e+04 | 1.2e+04 | 9501.13 |

Source: Authors' calculation based on PLFS Survey, 2020-21

Note: (i) *** signals significant at 1% level, ** signals significant at 5% level, and * signals significant at 10% level.

(ii) Standard errors are clustered at FSU levels for logit model.

(iii) The analysis controls for education categories, age-cohorts, gender, sector, social group, religion, and the ratio of eligible men to women in rural and urban areas, 36 states and union territories.

(iv) The Lewbel IV models contain three instrumental variables, namely, migrant network, crime, and unemployment rate in rural and urban areas. The first stage results are reported in the annexure (table A.12 for the reason of migration and Table A.13 for migration type).

**Table A.9: Heterogeneity Analysis: Regression Estimates for Educational Homogamy among Migrants: Based on Migration Stream**

| Dependent Variable: | (1) | (2) |
|---|---|---|
| Homogamous Union | Logit | IV |
| Base cat.: Non-migrant | | |
| Rural-rural | -0.101*** | -0.012* |
|  | (0.019) | (0.006) |
| Rural-urban | -0.133*** | -0.029*** |
|  | (0.028) | (0.006) |
| Urban-urban | -0.073** | 0.005 |
|  | (0.037) | (0.008) |
| Urban-rural | 0.053* | 0.002 |
|  | (0.029) | (0.006) |
| Observations | 128,400 | 125,996 |
| F-statistic |  | 1218.266 |
| Under-identification Test |  | 3.0e+04 |

Source: Authors' calculation based on PLFS Survey, 2020-21

Note: (i) *** signals significant at 1% level, ** signals significant at 5% level, and * signals significant at 10% level.
(ii) Standard errors are clustered at FSU levels for logit model.
(iii) The analysis controls for education categories, age-cohorts, gender, sector, social group, religion, and the ratio of eligible men to women in rural and urban areas, 36 states and union territories.
(iv) The Lewbel IV models contain three instrumental variables, namely, migrant network, crime, and unemployment rate in rural and urban areas. The first stage results are reported in the annexure (table A.14).

**Table A.10: Regression Estimates for Educational Homogamy among Migrants: Based on Reason of Migration (Gender-wise)**

|  | (1) | (2) | (3) | (1) |
|---|---|---|---|---|
| Dependent Variable: Homogamous Union | OLS | OLS | IV | IV |
|  | Male | Female | Male | Female |
| Base cat.: Non-migrant |  |  |  |  |
| Employment | -0.011 | 0.014 | -0.022** | 0.012 |
|  | (0.036) | (0.089) | (0.009) | (0.021) |
| Marriage | -0.177* | -0.077*** | -0.033 | -0.017*** |
|  | (0.107) | (0.022) | (0.020) | (0.005) |
| Other | 0.026 | -0.053 | -0.011 | -0.004 |
|  | (0.049) | (0.041) | (0.011) | (0.010) |
|  |  |  |  |  |
| Observations |  | 61,093 | 67,307 | 59,949 |

Source: Authors' calculation based on PLFS Survey, 2020-21

Note: (i) *** signals significant at 1% level, ** signals significant at 5% level, and * signals significant at 10% level.

(ii) Standard errors are clustered at FSU levels.

(iii) The analysis controls for education categories, age-cohorts, gender, sector, social group, religion, and the ratio of eligible men to women in rural and urban areas, 36 states and union territories.

(iv) The Lewbel IV models contain three instrumental variables, namely, migrant network, crime, and unemployment rate in rural and urban areas.

(v) The "other" reason for migration includes health-related reasons, housing problems, natural disasters, housing problems, migration of parent/another family member to pursue studies, displacement of development projects, acquisition of own house/flat, post-retirement, and others.

**Table A.11: IV Estimates: First stage estimates for Migrant Status**

| Dependent Variable: Migrant | (1) Migrant |
|---|---|
| Crime | -0.0323*** |
|  | (0.0024) |
| Unemployment rate (Rural) | 0.0022*** |
|  | (0.0008) |
| Unemployment rate (Urban) | -0.0041*** |
|  | (0.0006) |
| Migrant Network | 0.0356*** |
|  | (0.0032) |
|  |  |
| Observations | 125,996 |

Source: Authors' calculation based on PLFS Survey, 2020-21

Note: (i) *** signals significant at 1% level, ** signals significant at 5% level, and * signals significant at 10% level.

(ii) The analysis controls for education categories, age-cohorts, gender, sector, social group, religion, and eligible men-to-women ratio in rural and urban areas, 36 states and union territories.

**Table A.12: IV Estimates: First stage Estimates for Couple Migration Status**

|  | (1) | (2) | (3) |
|---|---|---|---|
| Dependent Variable: Couple Migration Status | Husband-migrant | Wife-migrant | Both Migrant |
| Crime | -0.0017*** | 0.0360*** | -0.0470*** |
|  | (0.0002) | (0.0018) | (0.0014) |
| Unemployment rate (Rural) | -0.0002*** | -0.0027*** | 0.0083*** |
|  | (0.0000) | (0.0005) | (0.0004) |
| Unemployment rate (Urban) | 0.0018*** | 0.0019*** | -0.0052*** |
|  | (0.0000) | (0.0003) | (0.0002) |
| Migrant Network | 0.0011*** | -0.0494*** | 0.0603*** |
|  | (0.0002) | (0.0023) | (0.0019) |
| Observations | 125,996 | 125,996 | 125,996 |

Source: Authors' calculation based on PLFS Survey, 2020-21

Note: (i) *** signals significant at 1% level, ** signals significant at 5% level, and * signals significant at 10% level.
(ii) The analysis controls for education categories, age-cohorts, gender, sector, social group, religion, eligible men to women ratio in rural and urban areas, 36 states and union territories.

Table A.13: IV Estimates: First stage Estimates for Reason of Migration

| Dependent Variable: Migration Type | (1) Employment | (2) Marriage | (3) Other |
|---|---|---|---|
| Crime | -0.0189*** | -0.0075*** | -0.0200*** |
|  | (0.0006) | (0.0011) | (0.0006) |
| Unemployment rate (Rural) | 0.0041*** | 0.0008** | 0.0039*** |
|  | (0.0001) | (0.0003) | (0.0001) |
| Unemployment rate (Urban) | -0.0026** | -0.0001 | -0.0025*** |
|  | (0.0001) | (0.0003) | (0.0014) |
| Migrant Network | 0.0225*** | 0.0046** | 0.0231*** |
|  | (0.0008) | (0.0014) | (0.0009) |
|  |  |  |  |
| Observations | 125,996 | 125,996 | 125,996 |

Source: Authors' calculation based on PLFS Survey, 2020-21

Note: (i) *** signals significant at 1% level, ** signals significant at 5% level, and * signals significant at 10% level.

(ii) The analysis controls for education categories, age-cohorts, gender, sector, social group, religion, ratio of eligible men to women in rural and urban areas, 36 states and union territories.

(iii) The "other" reason of migration includes health related reasons, housing problems, natural disaster, housing problems, migration of parent/other family member, pursue studies, displacement of development projects, acquisition of own house/flat, post-retirement, and others.

## Table A.14: IV Estimates: First stage Estimates for Migration Type

|  | (1) | (2) | (3) |
|---|---|---|---|
| Dependent Variable: Migration Type | Intra-district | Inter-district | Inter-state |
| Crime | 0.0064** | -0.0234*** | -0.0314*** |
|  | (0.0017) | (0.0012) | (0.0011) |
| Unemployment rate (Rural) | -0.0060*** | 0.0045*** | 0.0075*** |
|  | (0.0006) | (0.0004) | (0.0004) |
| Unemployment rate (Urban) | 0.0017** | -0.0028** | -0.0043*** |
|  | (0.0005) | (0.0003) | (0.0003) |
| Migrant Network | -0.0047 | 0.0184*** | 0.0337*** |
|  | (0.0030) | (0.0023) | (0.0029) |
|  |  |  |  |
| Observations | 125,996 | 125,996 | 125,996 |

Source: Authors' calculation based on PLFS Survey, 2020-21
Note: (i) *** signals significant at 1% level, ** signals significant at 5% level, and * signals significant at 10% level.
(ii) The analysis controls for education categories, age-cohorts, gender, sector, social group, religion, ratio of eligible men to women in rural and urban areas, 36 states and union territories.

Table A.15: IV Estimates: First stage Estimates for Migration Stream

| Dependent Variable: Migration Stream | (1) Rural-Rural | (2) Rural-Urban | (3) Urban-Urban | (1) Urban-Rural |
|---|---|---|---|---|
| Crime | 0.0113*** | -0.0022*** | 0.0007*** | -0.0021*** |
|  | (0.0014) | (0.0003) | (0.0001) | (0.0003) |
| Unemployment rate (Rural) | -0.0043*** | 0.0014*** | 0.0003 | -0.0006 |
|  | (0.0005) | (0.0001) | (0.0001) | (0.0001) |
| Unemployment rate (Urban) | 0.0017** | -0.0004*** | -0.0007 | 0.0002** |
|  | (0.0003) | (0.0008) | (0.0001) | (0.0001) |
| Migrant Network | -0.0137*** | 0.0027*** | -0.0005** | 0.0034*** |
|  | (0.0019) | (0.0004) | (0.0002) | (0.0004) |
| Observations | 125,996 | 125,996 | 125,996 | 125,996 |

Source: Authors' calculation based on PLFS Survey, 2020-21

Note: (i) *** signals significant at 1% level, ** signals significant at 5% level, and * signals significant at 10% level.

(ii) Standard errors are clustered at FSU levels.

(iii) The analysis controls for education categories, age-cohorts, gender, sector, social group, religion, ratio of eligible men to women in rural and urban areas, 36 states and union territories.

**Table A.16: Multinomial Regression Estimates for Educational Homogamy among Migrants: Based on Reason of Migration, Migration Type, and Migration Stream**

| Base Category: Homogamous Union | (1) Hypogamous | (2) Hypergamous |
|---|---|---|
| **Panel (a): Reason of Migration** | | |
| Base cat.: Non-migrant | | |
| Employment | -0.061 | 0.081** |
|  | (0.045) | (0.037) |
| Marriage | 0.092*** | 0.091*** |
|  | (0.024) | (0.020) |
| Other | -0.077* | 0.099** |
|  | (0.046) | (0.039) |
| Observations | 128,400 | 128,400 |
| **Panel (b): Migration Type** | | |
| Base cat.: Non-migrant | | |
| Intra-district | 0.026 | 0.092*** |
|  | (0.024) | (0.021) |
| Inter-district | 0.052* | 0.100*** |
|  | (0.031) | (0.026) |
| Inter-state | -0.005 | 0.057 |
|  | (0.050) | (0.042) |
| Observations | 128,400 | 128,400 |
| **Panel (c): Migration Stream** | | |
| Base cat.: Non-migrant | | |
| Rural-rural | 0.079*** | 0.105*** |
|  | (0.025) | (0.021) |
| Rural-urban | 0.004 | 0.200*** |
|  | (0.036) | (0.031) |
| Urban-urban | 0.104** | 0.057 |
|  | (0.048) | (0.041) |
| Urban-rural | -0.065* | -0.063* |
|  | (0.036) | (0.033) |
| Observations | 128,400 | 128,400 |

Source: Authors' calculation based on PLFS Survey, 2020-21

Note: (i) *** signals significant at 1% level, ** signals significant at 5% level, and * signals significant at 10% level.
(ii) Standard errors are clustered at FSU levels.
(iii) The analysis controls for education categories, age-cohorts, gender, sector, social group, religion, ratio of eligible men to women in rural and urban areas, 36 states and union territories.
(iv) The "other" reason of migration includes health related reasons, housing problems, natural disaster, housing problems, migration of parent/other family member, pursue studies, displacement of development projects, acquisition of own house/flat, post-retirement, and others.

**Table A.17: Diagnostic Tests for Lewbel IV Estimation**

| Test | (1) Migration | (2) Migration Type | (3) Migration Stream | (4) Couple Migration | (5) Migration Reason |
|---|---|---|---|---|---|
| Breusch–Pagan/Cook–Weisberg test for heteroskedasticity | 989.24 (0.000) | 988.94 (0.000) | 981.10 (0.000) | 1000.44 (0.000) | 990.36 (0.000) |
| F-test of excluded instruments | 216.92 | (i) 2715 (ii) 5080.14 (iii) 1956.14 | (i) 1662.46 (ii) 24292.98 (iii) 89324.41 (iv) 25946.83 | (i) 50915.76 (ii) 2661.11 (iii) 1389.71 | (i) 3613.82 (ii) 3214.53 (iii) 3596.92 |
| Under identification test (Kleibergen–Papp rk LM statistics) | 6382.303 (0.000) | 1.5e+04 (0.000) | 3.0e+04 (0.000) | 2.8e+04 (0.000) | 3.4e+04 (0.000) |
| Weak identification test (Cragg–Donald Wald F statistics) | 437.526 | 348.765 | 603.717 | 1365.288 | 1679.121 |
| Endogeneity test | 0.003 (0.9587) | 4.092 (0.2517) | 424.767 (0.000) | 21.417 (0.0001) | 27.108 (0.000) |
| Hansen J statistics | 121.829 (0.0000) | 341.115 (0.000) | 512.428 (0.000) | 316.764 (0.000) | 293.700 (0.000) |

Source: Authors' calculations using PLFS (2020-21)

Notes: (a) p-value is indicated in the parenthesis.
(b) In columns (1), (i), (ii), and (iii) indicate the f-test value of intra-district, inter-district, and inter-state, respectively.
(c) columns (2), (i), (ii), (iii), and (iv) indicate the f-test values of rural-rural, rural-urban, urban-urban, and urban-rural, respectively.
(d) Columns (3), (i), (ii), and (iii) indicate the f-test values of the husband-migrant, wife-migrant, and both migrants, respectively.
(e) In columns (4), (i), (ii), and (iii) indicate the f-test value of employment, marriage, and other reasons, respectively.

# Table A.18: Heckman Sample Selection: Probit Estimates for Family Structure

| Dependent Variable: Family Structure (Joint Family) | (1) Joint Family |
|---|---|
| Household Head Age | 0.058*** |
|  | (0.000) |
| Proportion of Durable Goods | 0.001*** |
|  | (0.000) |
| Share of Dependents | 0.017*** |
|  | (0.000) |
| Observations | 268,934 |

Source: Authors' calculation based on PLFS Survey, 2020-21

Note: (i) *** signals significant at 1% level, ** signals significant at 5% level, and * signals significant at 10% level.

(ii) The analysis also controls for education categories, age-cohorts, gender, sector, social group, religion, and eligible men-to-women ratio in rural and urban areas, 36 states and union territories.

Table A.19: Alternate Migration Network Variable: Regression Estimates for Educational Homogamy among Migrants

| Dependent Variable: | (1) | (2) | (3) |
|---|---|---|---|
| Educational Homogamy | IV | IV | IV |
|  |  | Rural | Urban |
| **Panel 1: Migrant Experience** | | | |
| Panel (1A): Migrant Status | | | |
| Migrant | -0.019** | -0.009 | -0.024* |
|  | (0.009) | (0.010) | (0.013) |
| Observations | 125,996 | 75,863 | 50,133 |
| Panel (1B): Couple Migration Status | | | |
| Base cat.: Non-migrant | | | |
| Husband Migrant | -0.021** | -0.004 | -0.037*** |
|  | (0.010) | (0.014) | (0.014) |
| Wife Migrant | -0.010*** | -0.005 | -0.020*** |
|  | (0.004) | (0.005) | (0.006) |
| Both Migrant | -0.025*** | -0.017* | -0.021*** |
|  | (0.006) | (0.009) | (0.007) |
| Observations | 125,996 | 75,863 | 50,133 |
| **Panel 2: Migrant Fractionalisation Index** | | | |
| Panel (2A): Migrant Status | | | |
| Migrant | -0.018** | -0.008 | -0.025* |
|  | (0.009) | (0.010) | (0.013) |
| Observations | 125,996 | 75,863 | 50,133 |
| Panel (2B): Couple Migration Status | | | |
| Base cat.: Non-migrant | | | |
| Husband Migrant | -0.022** | -0.005 | -0.037*** |
|  | (0.010) | (0.014) | (0.014) |
| Wife Migrant | -0.010*** | -0.005 | -0.020*** |
|  | (0.004) | (0.005) | (0.006) |
| Both Migrant | -0.025*** | -0.018* | -0.022*** |
|  | (0.006) | (0.009) | (0.007) |
| Observations | 125,996 | 75,863 | 50,133 |

Source: Authors' calculation based on PLFS Survey, 2020-21

Note: (i) *** signals significant at 1% level, ** signals significant at 5% level, and * signals significant at 10% level.

(ii) Standard errors are clustered at FSU levels.

(iii) The analysis controls for education categories, age-cohorts, gender, sector, social group, religion, and eligible men-to-women ratio in rural and urban areas, 36 states and union territories.

**Table A.20: Regression Estimates for Educational Homogamy among Migrants (with Sampling Weights)**

| Dependent Variable: Educational Homogamy | (1) Logit | (2) Logit | (3) Logit | (4) IV | (5) IV | (6) IV |
|---|---|---|---|---|---|---|
| | | Rural | Urban | | Rural | Urban |
| **Panel (a): Migrant Status** | | | | | | |
| Migrant | -0.075** | -0.072* | -0.040 | -0.018*** | -0.017*** | -0.009*** |
| | (0.031) | (0.043) | (0.039) | (0.000) | (0.000) | (0.000) |
| Observations | 367,628,356 | 270,957,106 | 96,671,250 | 361,263,948 | 265,879,410 | 95,384,538 |
| **Panel (b): Couple Migrant Status** | | | | | | |
| Base cat.: Non-migrant | | | | | | |
| Husband Migrant | -0.084 | 0.071 | -0.281** | -0.020*** | 0.015*** | -0.065*** |
| | (0.106) | (0.164) | (0.116) | (0.000) | (0.000) | (0.000) |
| Wife Migrant | -0.072* | -0.055 | -0.090* | -0.015*** | -0.013*** | -0.020*** |
| | (0.041) | (0.054) | (0.046) | (0.000) | (0.000) | (0.000) |
| Both Migrant | -0.098 | -0.143 | -0.058 | -0.025*** | -0.034*** | -0.013*** |
| | (0.060) | (0.099) | (0.066) | (0.000) | (0.000) | (0.000) |
| Observations | 367,628,356 | 270,957,106 | 96,671,250 | 361,263,948 | 265,879,410 | 95,384,538 |

Source: Authors' calculation based on PLFS Survey, 2020-21. Frequency weights are used.

Note: (i) *** signals significant at 1% level, ** signals significant at 5% level, and * signals significant at 10% level.

(ii) Standard errors are clustered at FSU levels for logit model.

(iii) The analysis controls for education categories, age-cohorts, gender, sector, social group, religion, and eligible men-to-women ratio in rural and urban areas, 36 states and union territories.

(iv) The Lewbel IV models contain three instrumental variables, namely, migrant network, crime, and unemployment rate in rural and urban areas. The first stage results are reported in the annexure (table A.10 for migrant status and table A.11 for couple migration status).

**Table A.21: Mechanism: Regression Estimates for Educational Homogamy: Based on Family Structure (Nuclear and Joint Family) (Excluding Borderline Cases)**

|  | (1) | (2) | (1) | (2) | (3) | (4) | (5) | (6) |
|---|---|---|---|---|---|---|---|---|
|  | Nuclear | Joint | Nuclear | Joint | Nuclear | | Joint | |
| Dependent Variable: Homogamous Union | Household Head's Age > 29 | | Household Head's Age < 50 | | Exclude top 10 and bottom 10 percentiles for Share of Durable household | | Exclude top 10 and bottom 10 percentiles for Share of Dependent households | |
| Base cat.: Non-migrant | | | | | | | | |
| Migrant | -0.049** | -0.095*** | -0.059** | -0.086* | -0.061** | -0.118*** | -0.051** | -0.093*** |
|  | (0.021) | (0.027) | (0.024) | (0.048) | (0.025) | (0.032) | (0.021) | (0.027) |
| Female | -0.069*** | -0.040** | -0.047*** | -0.086** | -0.078*** | -0.035 | -0.061*** | -0.040** |
|  | (0.015) | (0.019) | (0.017) | (0.035) | (0.018) | (0.023) | (0.014) | (0.019) |
| Observations | 75,154 | 46,672 | 62,554 | 15,408 | 54,764 | 33,951 | 80,884 | 47,345 |

Source: Authors' calculation based on PLFS Survey, 2020-21

Note: (i) *** signals significant at 1% level, ** signals significant at 5% level, and * signals significant at 10% level. (ii) Standard errors are clustered at FSU levels. (iii) The analysis controls for education categories, age-cohorts, gender, sector, social group, religion, and the ratio of eligible men to women in rural and urban areas, 36 states and union territories. The inverse mills ratio are also included. (iv) Columns (3), (4), (5), and (6) consist of multinominal regression where the base category of the dependent variable is educational homogamy.

**Table A.22: Regression Estimates for Educational Homogamy among Migrants (with unemployment rate using 2011-12 NSSO)**

|  | (4) | (5) | (6) |
|---|---|---|---|
| Dependent Variable: Educational Homogamy | IV | IV | IV |
|  |  | Rural | Urban |
| Migrant | -0.014*** | -0.012*** | -0.007*** |
|  | (0.000) | (0.000) | (0.000) |
| Observations | 277,435,157 | 203,539,983 | 73,895,174 |

Source: Authors' calculation based on PLFS Survey, 2020-21

Note: (i) *** signals significant at 1% level, ** signals significant at 5% level, and * signals significant at 10% level.
(ii) Standard errors are clustered at FSU levels for logit model.
(iii) The analysis controls for education categories, age-cohorts, gender, sector, social group, religion, and eligible men-to-women ratio in rural and urban areas, 36 states and union territories.
(iv) The Lewbel IV models contain three instrumental variables, namely, migrant network, crime, and unemployment rate in rural and urban areas (2011-12).

A.3 Annexure Figures

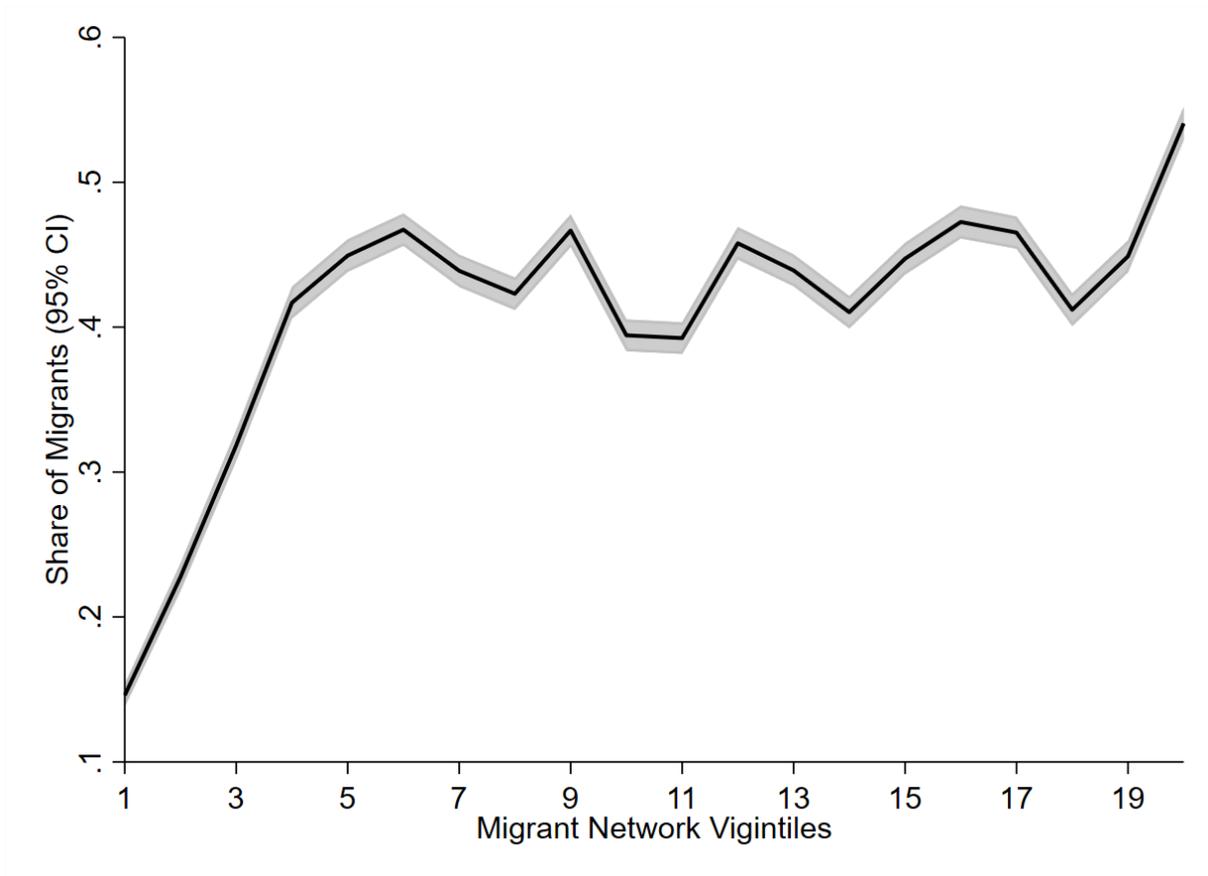

Figure A.1: Distribution of Migration Rates Across Migrant Network Vigintiles